\documentclass[nolineno]{jfm}

\usepackage{graphicx}
\usepackage{newtxtext}
\usepackage{newtxmath}
\usepackage{natbib}
\usepackage{hyperref}
\hypersetup{
    colorlinks = true,
    urlcolor   = blue,
    citecolor  = blue
}

\newcommand{\RomanNumeralCaps}[1]
\linenumbers

\newcommand\redout{\bgroup\markoverwith{\textcolor{red}{\rule[0.5ex]{2pt}{0.8pt}}}\ULon}
\renewcommand\emph[1]{{\it #1}}
\usepackage[super]{nth}

\input{config2.tex}
\usepackage{tikz}

\newcommand\dd{{\rm d}}

\title{Closely spaced co-rotating helical vortices: Long-wave instability}
\author{A. Castillo-Castellanos\aff{1}
  \corresp{\email{andres-alonso.castillo-castellanos@univ-amu.fr}},
 \and S. Le Diz\`es\aff{1}}

\affiliation{\aff{1}Aix Marseille Universit\'e, CNRS, Centrale Marseille, IRPHE, Marseille, France}

\begin{document}
\maketitle

\begin{abstract}
  We consider as base flow the stationary vortex filament solution obtained by
  \cite{castillo2020} in the far-wake of a rotor with tip-splitting blades. The
  cases of a single blade and of two blades with a hub vortex are studied. In
  these solutions, each blade generates two closely spaced co-rotating tip
  vortices that form a braided helical pattern in the far-wake. The long-wave
  stability of these solutions is analysed using the same vortex filament
  framework. Both the linear spectrum and the linear impulse response are
  considered. We demonstrate the existence of different types of instability
  modes. A first type corresponds to the local pairing of consecutive turns of
  the helical pattern, which is well described by the instability of an
  uniform helical vortex with a core size given by the mean separation distance
  of the vortices in the pair. {A second type corresponds to the pairing of
  consecutive turns of vortex pair and is observed only for densely braided
  patterns, which is well described by the instability of two interlaced helical
  vortices by straightening out the baseline helix. A third type of unstable
  modes modifies the separation distance between the vortices in each pair and
  amplifies specific (linear) wavelengths.} These unstable modes also
  spread spatially with a weaker rate.
\end{abstract}

\section{Introduction}

Rotating blades, such as those of a helicopter rotor or a horizontal-axis wind
turbine, generate concentrated vortices at their tips, transported downstream,
creating a distinctive helical pattern. These helical vortices are associated
with several practical issues actively investigated. One of these issues is the
wake stability with respect to external disturbances, which has practical
relevance since instabilities may accelerate the vortex break-up and the
transition towards turbulent wakes. The interaction between tip vortices and a
solid surface, like a trailing rotor blade, or a wind turbine located
downstream, may cause significant noise, vibration, and fatigue problems. Wake
stability is also fundamental to optimize wind farms design, since wakes may
extend up to 50 km downstream under stable atmospheric conditions
\citep{canadillas2020offshore}. Strategies to maximize power generation include
dynamically varying the yaw, pitch, and tip-speed ratio to excite the natural
instability of helical wakes (see, for instance \cite{huang2019numerical,
frederik2020helix, brown2022accelerated}).

One alternative to accelerate the vortex breakdown is to modify the wake
structure altogether. \cite{brocklehurst1994reduction} introduced a modified
vane tip to split the tip vortex into two co-rotating vortices. The associated
spreading of vorticity has been suggested as indicative of reduced noise and
increased aerodynamic efficiency, but there is little information regarding the
change in the wake structure, which is essential to optimize the air-foil
design. The present work aims to fill in some of these gaps by focusing on the
instabilities of closely-spaced helical vortices. A recent experimental
investigation of two closely spaced helical vortices generated by single-bladed
rotor, \cite{schroder2020experimental, schroder2021instability} displays a rapid
increase of the core radii and subsequent merging related to the development of
a centrifugal instability \citep{bayly1988three}. In this case, the centrifugal
instability is triggered by patches of opposite signed vorticity formed by a
protruded fin during the roll-up process. A theoretical analysis by
\cite{castillo2020} presented the wake geometry produced by a tip-splitting
rotor for all wind turbine and helicopter flight regimes using a filament
approach. The resulting wake deviates from a helical pattern in favour of an
epicycloidal pattern produced by two interlaced helical vortices inscribed on
top of a larger helical curve. From afar, the vortex structure is reminiscent of
a helical vortex, while simultaneously resembling a vortex pair aligned with the
locally tangent flow. Given these similarities, we expect the solutions to
display features from both systems. In particular, similar instability
mechanisms.

Studies on the stability of helical vortices started a century ago \citep[see
for instance][]{Levy1928} although experimental evidence of short-wave and long
wave instabilities are quite recent \citep{Leweke14}. Theoretical predictions
for the long-wave instability have been obtained by \cite{widnall1972stability}
for a uniform helical vortex. Her work was extended by
\cite{gupta1974theoretical} for several helices and by \cite{Fukumoto1991} to
account the effect of an axial flow within the vortices. As demonstrated by
\cite{quaranta2015long,quaranta2019local}, the long-wave instability is a local
pairing instability \citep{lamb1945hydrodynamics}. The instability modes are
associated with a displacement approaching neighbouring loops at
specific locations. \cite{quaranta2019local}  have also shown that their growth
rate is well-predicted by considering 3D perturbations on straight vortices
\citep{Robinson1982}. For several helices, additional theoretical results have
been obtained for the global pairing mode which preserves the helical symmetry
of the flow \citep{okulov2004stability,Okulov07} using Hardin's expressions for
the induced velocity of a helical vortex \citep{kawada1936induced,
hardin1982velocity}. This mode has been further analysed in the non-linear regime
by \cite{selccuk2017helical}. Finally, recent numerical works have also
evidenced the presence of the long-wave instability in the context of
rotor-generated vortices \citep{Bhagwat00, Walther07, ivanell2010stability,
venegas2020}. Helical vortices are also unstable to short-wavelength
instabilities, due to the modification of the core structure by curvature,
torsion and strain \citep{kerswell2002elliptical, blanco2016elliptic,
blanco2017curvature}. This requires detailed information of the inner core
structure, which can be obtained through matched asymptotic techniques
\citep{blanco2015internal}, or through direct numerical simulations
\citep{selccuk2017helical, brynjell2020numerical}. Pairs display a variety of
complex behaviors \citep{leweke2016dynamics}. For instance, consider a vortex
pair of circulations $\Gamma_1$ and $\Gamma_2$ and core radii $a$, separated by
a distance $d$. If $d$ is large enough, the system is expected to translate with
constant speed for $\Gamma_1=-\Gamma_2$, or rotate around the vorticity
barycentre with constant rotation rate for all other cases, while $a$ is
expected to grow following a viscous law. Pairs are expected to merge as a
critical core size is eventually reached \citep{meunier2002merging,
josserand2007merging}. This behaviour is modified by the development of
short-wave instabilities \citep{meunier2005elliptic} and long-wave instabilities
\citep{crow1970stability, jimenez1975stability}. This picture becomes
increasingly complex as we consider the interactions between multiple vortex
pairs, like the four vortex system considered by \citep{crouch1997instability,
fabre2000stability, fabre2002optimal}.

For this work, we focus on the long-wavelength stability of two closely-spaced
helical vortices using a filament approach. This approach has the advantage of
filtering the short-wavelength perturbations. Additionally, we are interested in
the regime observed in the far-field, since perturbations are expected to be
quickly advected away from the rotor \citep{venegas2020}. We shall use as base
flow the steady solutions obtained in \cite{castillo2020}.
{%
Depending on the geometric
parameters, these solutions are classified as {\it(i)} leapfrogging, {\it (ii)}
sparsely braided, and {\it (iii)} densely braided wakes.
From afar, the periodic structure is reminiscent of a helical vortex but
up close, it resembles two interlaced helical vortices aligned with the locally
tangent flow. Given these similarities, we expect he stability of the present
system to be understood as a combination of the pairing modes of
\begin{itemize}
  \item a helical vortex of radius $R$, pitch $H$, circulation $2\Gamma$, and
  some effective core size $a_e$
  \item a pair of helical vortices of radius $d/2$, pitch $h_\tau$,
  core size $a$, and circulation $\Gamma$
\end{itemize}
and by the possible interactions between them. We shall see that both kinds of
pairing are observed, as well as a new kind of unstable modes specific to this
configuration.
}

The paper is organized as follows. In section \ref{sec:background} we present
the framework of the vortex method applied to the pair of helical vortices. We
describe the numerical procedures to obtain the base flow and analyse its
stability. In section \ref{sec3}, we apply our numerical approach to uniform
helices for validation. We then consider two different configurations. The
stability properties for a pair of helical vortices without a central vortex hub
are presented in section \ref{sec4} {for the different wake geometries}, while the
stability properties for two pairs of helical vortices with a central hub vortex
are presented in section \ref{sec5} {for leapfrogging wakes}. In both cases, the
structure of the unstable modes is presented in detail and we explore the
influence of the main geometric parameters. Additional information regarding the
spatio-temporal development of the instability is provided in the appendix
\ref{secA}, where the linear impulse response for the first configuration is
studied using the approach developed by \cite{venegas2020}. Finally, we present
our main conclusions in section \ref{sec6}.

\section{Methodology}\label{sec:background}

\subsection{Finding stationary solutions using a filament approach}

\subsubsection{Filament framework}

We use the same vortex filament approach as in \cite{venegas2019generalized}.
Vorticity is considered to be concentrated along slender vortex filaments
moving as material lines in the fluid according to
\begin{equation}\label{sec2:eq-1}
  \frac{\dd \boldsymbol{X}_j}{\dd t} =
  \boldsymbol{U}(\boldsymbol{X}_j) - {U}_z^{\infty}\boldsymbol{\hat{e}_z}
\end{equation}
where $\boldsymbol{X}_j=(r_j,\theta_j,z_j)$ is the position of the $j$-th vortex
filament, ${U}_z^{\infty}$ an external velocity field, and
$\boldsymbol{U}(\boldsymbol{X}_j)=(U_r , r_j \Omega, U_z)$ is the velocity
induced by the vortices at $\boldsymbol{X}_j$. In a frame with rotation rate
$\Omega_R$, it is convenient to parametrize the position of vortex filaments in
terms of two angular coordinates, $\zeta=\Omega_R (t-t_0)$ and $\psi = \Omega_R
t$, and transform \eqref{sec2:eq-1} as
\begin{equation}\label{sec2:eq0}
  \frac{\partial \boldsymbol{X}_j}{\partial \psi} + \frac{\partial \boldsymbol{X}_j}{\partial \zeta} =
  \frac{1}{\Omega_R}\left[ \boldsymbol{U}(\boldsymbol{X}_j) + \boldsymbol{U}^{\infty}(\boldsymbol{X}_j) \right]
\end{equation}
where $\boldsymbol{U}^{\infty}(\boldsymbol{X}_j)=(0,-r_j \Omega_R,
-{U}_z^{\infty})$ contains the contributions from the rotating frame and a
constant free-stream velocity. In this context, $\zeta$ is often referred to as
the wake age and corresponds to a spatial coordinate, while $\psi$ is a proxy of
time \citep{leishman2002free}. Filaments are discretized in straight segments
$[\boldsymbol{X}^{n}_j, \boldsymbol{X}^{n+1}_j]$ in order to compute the
velocity field using the Biot-Savart law. The equation is de-singularized using a
cut-off approach with a Gaussian vorticity profile. Local contributions to the
velocity field at $\boldsymbol{X}_j^n$ are obtained by replacing the straight
segments by an arc of circle passing through $[\boldsymbol{X}^{n-1}_j,
\boldsymbol{X}^{n}_j, \boldsymbol{X}^{n+1}_j]$ and use the cut-off formula (see,
\cite{venegas2019generalized}).

For this work, we consider basic flow solutions that satisfy,
\begin{equation}\label{sec2:eq2}
  \frac{\dd r_j}{\dd\zeta} = \frac{U_r(\boldsymbol{X}_j)}{\Omega_R}, \quad
  \frac{\dd \theta_j}{\dd\zeta} = \frac{\Omega (\boldsymbol{X}_j)-\Omega_R}{\Omega_R}, \quad
  \frac{\dd z_j}{\dd\zeta} = \frac{U_z(\boldsymbol{X}_j) - U^\infty_z}{\Omega_R}.
\end{equation}
These solutions are $\psi$-independent solutions to \eqref{sec2:eq0}. They are
stationary in the moving frame in the sense that flow remains tangent to the
vortex structure at all times:
\begin{equation}\label{sec2:eq1}
  (\boldsymbol{U}(\boldsymbol{X}_j) + \boldsymbol{U}^{\infty}(\boldsymbol{X}_j)) \times \boldsymbol{T}_j = 0
\end{equation}
where $\boldsymbol{T}_j$ is the tangent unit vector. As shown in
\cite{castillo2020}, the spatial evolution of the wake as we move away from the
rotor plane is obtained by numerically solving \eqref{sec2:eq2} with
boundary conditions on the rotor, at $\zeta=0$, where the position of each
vortex is prescribed and far-field boundary conditions at $\zeta\to\infty$. In
the following, we focus on the wake geometry in the far-field, where
spatially periodic solutions are obtained.

\subsubsection{Periodic solutions in the far-field}

\begin{figure}\centering
  \includegraphics[width=0.9\linewidth, trim=0 0 0 0, clip]{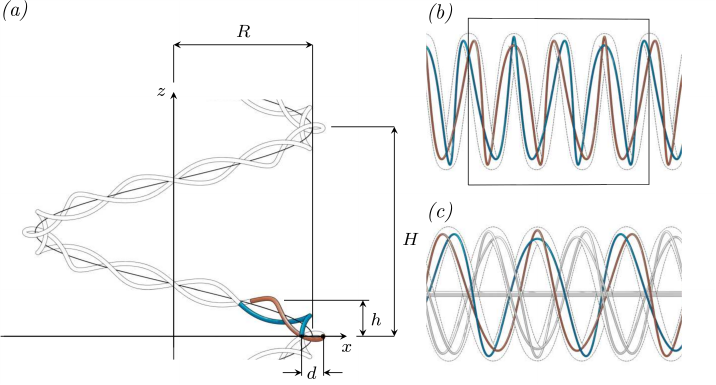}
  \caption{
  {\it (a)} Geometric parameters of periodic solutions: separation distance $d$,
  radius $R$, axial pitches $H$ and $h$, core size $a$ and $\beta = H/h$.
  Representation of {\it (b)} $N=1$ pairs without hub vortex and
  {\it (c)} $N=2$ pairs with hub vortex for ($R^*=7$, $H^*/N=5.25$, $\beta=3/4$).
  }
  \label{newfig:1}
\end{figure}

The solution in the far-field has been obtained in \cite{castillo2020}. It was
shown that it is no longer uniform as for a single helix but it nevertheless
exhibits a certain spatial periodicity. In addition to the azimuthal symmetry
$\theta \to \theta + 2 \pi /N$ for $N$ vortex pairs, solutions are invariant by
the double operation $z \to z + h$ and $\theta \to \theta + 2\pi/\beta$. The
parameters $h$ and $\beta$ are chosen such that  there is a single location in
an axial period $h$ where the vortices  of a given pair are at the same azimuth.
This azimuth is taken to define the mean radius $R$ and the separation distance
$d$ of each vortex pair. Each vortex pair tends to form a helical braid on a
larger helical structure of radius $R$ and pitch $H$ as illustrated on figure
\ref{newfig:1}.
%%%%%%%%%%%%%%%%%%%%%%%%%%%%%%%%%%%%%%%%%%%%%%%%%%%%%%%%%%%%%%%%%%%%%%%%%%%%%%%%
%%%%%%%%%%%%%%%%%%%%%%%%%%%%%%%%%%%%%%%%%%%%%%%%%%%%%%%%%%%%%%%%%%%%%%%%%%%%%%%%
%%%%%%%%%%%%%%%%%%%%%%%%%%%%%%%%%%%%%%%%%%%%%%%%%%%%%%%%%%%%%%%%%%%%%%%%%%%%%%%%
%%%%%%%%%%%%%%%%%%%%%%%%%%%%%%%%%%%%%%%%%%%%%%%%%%%%%%%%%%%%%%%%%%%%%%%%%%%%%%%%
The pitch of this larger helical structure is directly related to $h$ and $\beta$ by
\begin{equation}\label{exp:H}
  H= h \beta
\end{equation}
{%
and related to the pitch of the vortex pair through
\begin{equation}\label{exp:htau}
  h_\tau = \sqrt{H^2 + (2\pi R)^2}/\beta.
\end{equation}
}

{
Depending on the value of $\beta$, the double-helix structure may describe {\it
(i)} a leapfrog-type pattern, where vortices trade places every $1/\beta$ turns;
{\it (ii)} a relatively sparse braid ; or {\it (iii)} a dense `telephone
cord'-type pattern. We are typically in situation {\it (i)} when $\beta < 1$,
and in situation {\it (iii)} when $\beta > 10$.} The parameter $\beta$ also
characterizes the axial periodicity of the solution. It becomes axially periodic
only if $\beta$ is a rational number $p/q$. In that case, it means that the
double helix makes $q$ turns on itself as the large helix does $p$ turns. The
axial period is thus $pH=qh$. In the following, we only consider rational values
of $\beta$.

As soon as the vortex core size $a$ is fixed, the far-field is then defined by 5
non-dimensional parameters which are
 \begin{equation}\label{exp:parameters}
  R^* \equiv \frac{R}{d}, \quad
  H^* \equiv \frac{H}{d}, \quad
  \varepsilon^* \equiv \frac{a}{d}, \quad \beta, \quad N.
 \end{equation}

The solution is obtained by solving the system \eqref{sec2:eq2} with the
prescribed symmetry. The frame velocities $(\Omega_R,U^{\infty}_z)$ are unknown
quantities. However, these quantities are proportional to the vortex circulation
$\Gamma$ as it is the unique quantity of the vortex system involving time. For
this reason, we can fix $\Gamma$ to 1.

As shown in \cite{castillo2020}, the problem can be treated as a non-linear
minimization problem using an iterative procedure. The convergence is rapid if
we start for each pair from an initial guess given by an undeformed double helix
on a larger helix and estimates for $(\Omega_R,U^{\infty}_z)$ obtained from
uniform helices. The converged solution is found to exhibit spatial variations
but in most cases the initial guess turns out to be a good approximation of the
solution.

In the present study, we consider two different configurations: one composed of
a single helical pair ($N=1$) without central hub vortex (figure \ref{newfig:1}b), and
another composed of $N=2$ helical pairs with a central hub vortex (figure
\ref{newfig:1}c). Also, we fix $\varepsilon^*=0.1$ and vary the remaining
parameters ($R^*$, $H^*$ and $\beta$). In the following, solutions that satisfy
\eqref{sec2:eq2} are denoted $\boldsymbol{X}_j^B(\zeta)$.

\subsection{Inviscid global instability analysis}
\label{subsec:linear}
The stability of $\boldsymbol{X}_j^B$ is analysed by considering
the evolution of infinitesimal perturbation displacements
\begin{equation}
  \boldsymbol{X}_j(\zeta,\psi) = \boldsymbol{X}_j^B(\zeta) + \boldsymbol{X}'_j(\zeta,\psi)
\end{equation}
Equation \eqref{sec2:eq0} is linearized around $\boldsymbol{X}_j^B(\zeta)$ to
obtain a linear dynamical system
\begin{subequations}\label{sec:eqlinear}
\begin{eqnarray}
  \frac{\partial r_i'}{\partial\psi} &=& -\frac{\partial r_i'}{\partial\zeta}
  + \frac{1}{\Omega_R}\sum_{j=1}^N\left[
    \frac{\partial U_r(\vec{X}_i^B) }{\partial r_j} r_j' +
    \frac{\partial U_r(\vec{X}_i^B) }{\partial \theta_j} \theta'_j +
    \frac{\partial U_r(\vec{X}_i^B) }{\partial z_j} z'_j
    \right]
    \\
  \frac{\partial \theta_i'}{\partial\psi} &=& -\frac{\partial \theta_i'}{\partial\zeta}
  + \frac{1}{\Omega_R}\sum_{j=1}^N\left[
    \frac{\partial \Omega(\vec{X}_i^B) }{\partial r_j} r_j' +
    \frac{\partial \Omega(\vec{X}_i^B) }{\partial \theta_j} \theta'_j +
    \frac{\partial \Omega(\vec{X}_i^B) }{\partial z_j} z'_j
    \right]
    \\
  \frac{\partial z_i'}{\partial\psi} &=& -\frac{\partial z_i'}{\partial\zeta}
  + \frac{1}{\Omega_R}\sum_{j=1}^N\left[
    \frac{\partial U_z(\vec{X}_i^B) }{\partial r_j} r_j' +
    \frac{\partial U_z(\vec{X}_i^B) }{\partial \theta_j} \theta'_j +
    \frac{\partial U_z(\vec{X}_i^B) }{\partial z_j} z'_j
    \right]
\end{eqnarray}
\end{subequations}
which has the following general form
\begin{equation}\label{sec2:eq14}
  \frac{\partial\boldsymbol{q}'}{\partial\psi} = \tensor{L}(\zeta)\boldsymbol{q}'(\zeta, \psi)
\end{equation}
where $\boldsymbol{q}'=(\boldsymbol{X}_1, \boldsymbol{X}_2, \cdots)$ is the
total displacement vector, and $\tensor{L}$ is a $N$ by $N$ block-matrix
containing the Jacobian terms. The spatial derivative (on $\zeta$) is evaluated
using the same finite differences scheme as the base solution, while the
velocity gradient is derived from the discretized Biot-Savart equations.
%{see appendix \ref{secB}}.

\begin{figure}\it
  \hspace{0.5cm} (a) \hfill (b) \hfill ~ \\
  \includegraphics[width=0.95\linewidth]{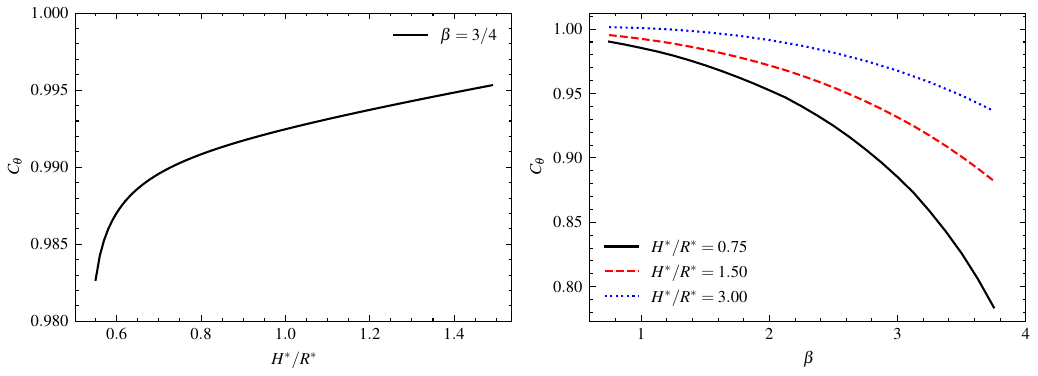}
  \caption{
    Evolution of the ratio $C_\theta\equiv\zeta_B/\theta_j(\zeta_B)$ for ($N=1$,
    $R^*=7$): {\it (a)} $C_\theta$ as function $H^*/R^*$ for $\beta=3/4$, and
    {\it (b)} as function of $\beta$ for different $H^*/R^*$.
  }
  \label{newfig:2}
\end{figure}

Each sub-matrix of $\tensor{L}$ can be written as
\begin{equation}
  \tensor{L}_{ij} =
  -\delta_{ij}\frac{\partial}{\partial \zeta} +
  \frac{1}{\Omega_R}
  \frac{\partial(\boldsymbol{U}(\boldsymbol{X}_i^B))}{\partial(\boldsymbol{X}_j)}
\end{equation}
and contains $3n_s$ by $3n_s$ elements where $n_s$ is the total number of
discretization point of each vortex.
The base flow is periodic with respect to $\zeta$ with a period
$\zeta_B$ that satisfies $z_j(\zeta_B) = h$ and $\theta_j(\zeta_B)=2\pi/\beta$.
The difference between $\zeta_B$ and $\theta_j(\zeta_B)$ comes from the induced
angular velocity $\Omega(\boldsymbol{X}_j)$ in \eqref{sec2:eq2}b. For small
values of $\beta$, the ratio $C_\theta \equiv \zeta_B/\theta_j(\zeta_B)$ is
close to 1, but decreases as $\beta$ increases (figures \ref{newfig:2}a and
\ref{newfig:2}b).
Each vortex contains 96 points per period $\zeta_B$. For the perturbations we
consider a longer domain with a period $\zeta_p = n_p \zeta_B $ with $n_p =
32\beta$ such that $z_j(\zeta_p) = n_p h = 32 H$ and $\theta_j(\zeta_p)= n_p
2\pi/\beta = 64\pi $. This gives $n_p=24$ and $n_p=256$ for the values
$\beta=3/4$ and $\beta=8$.
{%
For densely braided wakes ($\beta>20$), we use a shorter domain with $n_p =
4\beta$, such that $z_j(\zeta_p) = n_p h = 4 H$ and $\theta_j(\zeta_p)= n_p
2\pi/\beta = 8\pi $. This gives $n_p=84$ and $n_p=128$ for the values
$\beta=21$ and $\beta=32$.
}
Our objective is to apply periodic boundary conditions to the perturbations. We
therefore discretize the operators $\tensor{L}_{ij}$ such that they satisfy this
property.

%%%%%%%%%%%%%%%%%%%%%%%%%%%%%%%%%%%%%%%%%%%%%%%%%%%%%%%%%%%%%%%%%%%%%%%%%%%%%%%%
%%%%%%%%%%%%%%%%%%%%%%%%%%%%%%%%%%%%%%%%%%%%%%%%%%%%%%%%%%%%%%%%%%%%%%%%%%%%%%%%
%%%%%%%%%%%%%%%%%%%%%%%%%%%%%%%%%%%%%%%%%%%%%%%%%%%%%%%%%%%%%%%%%%%%%%%%%%%%%%%%
%%%%%%%%%%%%%%%%%%%%%%%%%%%%%%%%%%%%%%%%%%%%%%%%%%%%%%%%%%%%%%%%%%%%%%%%%%%%%%%%

The final problem reduces to a linear system with constant coefficients that
admits the formal solution
\begin{equation}\label{sec2:eq16}
  \boldsymbol{q}'(\zeta,\psi) = \exp(\tensor{L}\psi)\boldsymbol{q}_0'(\zeta)
  \end{equation}
for any initial perturbation
$\boldsymbol{q}'_0(\zeta)=\boldsymbol{q}'(\zeta,0)$. In standard fashion,
treating equation \eqref{sec2:eq16} as an eigenvalue problem can be used to
describe the asymptotic limit $\psi\to\infty$. Here, because the domain is long
but periodic, it also means that we consider perturbations that span the whole
calculation domain.

We may decompose the linear operator as
\begin{equation}
  \tensor{L} \tensor{\Phi} = \tensor{\Phi}\tensor{\Lambda}
\end{equation}
where $\tensor{\Phi}$ is a matrix whose columns $(\boldsymbol\phi_1,
\boldsymbol\phi_2, \cdots)$ are the eigenvectors of $\tensor{L}$, and
$\tensor{\Lambda}$ is a matrix whose diagonal entries are the corresponding
(complex) eigenvalues $(\lambda_1, \lambda_2, \cdots)$. Here, $\sigma_m \equiv
\Real(\lambda_m)$ characterizes the temporal growth (or decay) of the
perturbations, while $\omega_m\equiv\Imag(\lambda_m)$ characterizes the temporal
oscillations. We may also introduce a dimensionless growth rate $\sigma_m^*$ and
frequency $\omega_m^*$, based on the characteristic advection time-scale of
helical pairs $t_{adv} = \Gamma/(H^2/N^2)$. Since $\tensor{L}$ is real-valued,
modes are either real or come in conjugate pairs. Equation \eqref{sec2:eq16} can
be written as
\begin{equation}\label{sec2:eq17}
  \boldsymbol{q}'(\zeta,\psi) = \sum_m \boldsymbol{\phi}_m (\zeta)\exp{\lbrace\sigma_m\psi + i \omega_m \psi\rbrace} b_m + c.c.
\end{equation}
where the coefficients $b_m$ correspond to the coordinates of
$\boldsymbol{q}'_0(\zeta) = \boldsymbol{q}'(\zeta,0)$ expressed in the
eigenvector basis, and $c.c.$ indicates the complex conjugate.

By construction, eigenvectors have the same dimensionality as
$\boldsymbol{q'}$,
\begin{equation}
  \boldsymbol{\phi}_m = (\boldsymbol\phi_m^{(1)}, \boldsymbol\phi_{m}^{(2)}, \cdots)
\end{equation}
where $\boldsymbol\phi_{m}^{(j)}$ is the $m$-th eigenvector of the $j$-th vortex
filament. These eigenvectors represent a displacement vector and can be
expressed in any coordinate system, e.g.,
\begin{equation}
  \boldsymbol\phi_{m}^{(j)} = (\check{r}_{m}^{(j)},\check{\theta}_{m}^{(j)},\check{z}_{m}^{(j)})
\end{equation}
indicates the components of $\boldsymbol\phi_{m}^{(j)}$ in global cylindrical
coordinates. Conversely, displacement perturbations may also be expressed in
terms of the local radial, azimuthal, and axial coordinates associated with a
uniform helix $\mathscr{H}$ of radius $R$ and pitch $H$,
\begin{equation}
  \boldsymbol\phi_{m}^{(j)} = (\check{\rho}_{m}^{(j)}, \check{\phi}_{m}^{(j)}, \check{s}_{m}^{(j)}).
\end{equation}

Due to the spatial periodicity, the eigenvectors can be expanded on a discrete
Fourier basis:
\begin{equation}
  \boldsymbol{\phi}_m
  = \sum_{n=0}^{n_s-1} \boldsymbol{\hat{q}}_{mn} \exp{\left\lbrace \frac{\mathrm{i} 2\pi n\zeta}{\zeta_p} \right\rbrace}
\end{equation}
where the azimuthal wavenumber $(k = n \Delta k, \Delta k = \beta/n_p)$ is
normalized to ensure that $k=1$ corresponds to one helix turn, and
$n_s$ the number of points used to discretize each vortex.

\section{Validation of the stability analysis: uniform helices}
\label{sec3}

We validate our approach against existing results on uniform helical vortices of
pitch $H$, radius $R$ and effective core size $a_e$. A notable difference with
respect to \cite{widnall1972stability} is that {we must specify a reference frame
with rotation rate $\Omega_R$ such that \eqref{sec2:eq2} is satisfied}. Since the
motion of helical vortices is described by constant rotation rate $\Omega$ and
axial velocity $U_z$, these vortices are unperturbed by an additional rotation
of angular velocity $\Omega^a$ and translation of axial velocity $U_z^a$
provided that \begin{equation} \frac{U_z^a}{\Omega^a} = \pm \frac{H}{2\pi}
\end{equation} is satisfied, where the sign is positive for right-handed helices
and {\it vice versa}. For this comparison, we consider a rotating frame of
reference
\begin{equation}
  \Omega_R = \Omega \mp U_z \frac{2\pi}{H}, \quad\quad
  U_z^\infty = 0
\end{equation}
which is commonly used in numerical simulations.

%%%%%%%%%%%%%%%%%%%%%%%%%%%%%%%%%%%%%%%%%%%%%%%%%%%%%%%%%%%%%%%%%%%%%%%%%%%%%%%%
%%%%%%%%%%%%%%%%%%%%%%%%%%%%%%%%%%%%%%%%%%%%%%%%%%%%%%%%%%%%%%%%%%%%%%%%%%%%%%%%
\begin{figure}\it \centering
  \hspace{0.5cm} (a) \hfill (b) \hfill~ \\
  \includegraphics[width=0.95\linewidth, trim=0 190 0 0, clip]{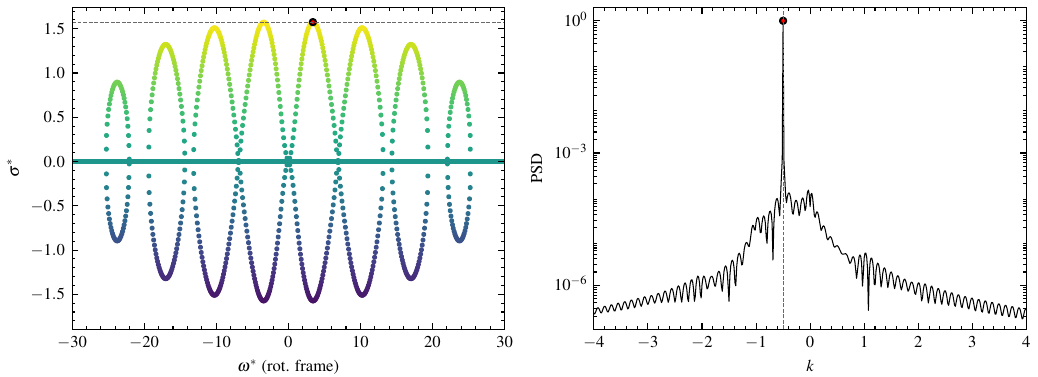}

  \hspace{0.5cm} (c) \hfill (d) \hfill~ \\
  \includegraphics[width=0.95\linewidth, trim=0 0 0 185, clip]{./new_figures/figure3.pdf}
  \caption{
    Stability of a uniform helix for $H/R=\pi/5$ and $a_e/R=0.1$: {\it (a)}
    $\sigma^*$ as function of $\omega^*$ in the rotating frame; {\it (b)}
    Fourier spectra of the mode indicated as a red mark in {\it (a)}; {\it
    (c)} $\sigma^*$ and {\it (d)} $\omega^*$ in the advection frame as function of
    dominant wavenumber $k_0$.
  }
  \label{newfig:3}
\end{figure}
%%%%%%%%%%%%%%%%%%%%%%%%%%%%%%%%%%%%%%%%%%%%%%%%%%%%%%%%%%%%%%%%%%%%%%%%%%%%%%%%
%%%%%%%%%%%%%%%%%%%%%%%%%%%%%%%%%%%%%%%%%%%%%%%%%%%%%%%%%%%%%%%%%%%%%%%%%%%%%%%%

%%%%%%%%%%%%%%%%%%%%%%%%%%%%%%%%%%%%%%%%%%%%%%%%%%%%%%%%%%%%%%%%%%%%%%%%%%%%%%%%
%%%%%%%%%%%%%%%%%%%%%%%%%%%%%%%%%%%%%%%%%%%%%%%%%%%%%%%%%%%%%%%%%%%%%%%%%%%%%%%%
\begin{table}\centering
  \begin{tabular}{c|rrr|rrr|rrr|}
               &\multicolumn{3}{c|}{$a_e/R=0.10$} & \multicolumn{3}{c|}{$a_e/R=0.20$} & \multicolumn{3}{c|}{$a_e/R=0.33$} \\
               & \nth{1} Peak & \nth{2} Peak & \nth{3} Peak & \nth{1} Peak & \nth{2} Peak & \nth{3} Peak & \nth{1} Peak & \nth{2} Peak & \nth{3} Peak \\
    $\sigma^*$ &  1.575 & 1.511 & 1.323 &  1.554 & 1.497  &  1.381 &  1.538 &  1.485 &  1.395 \\
    $\omega^*$ &  3.405 & 10.202 & 16.980 &  3.186 & 9.639  &  15.985 &  3.028 &  9.160 &  15.190 \\
    $k_0$      & -0.500 &-1.500 &-2.500 & -0.500 & -1.516 & -2.516 & -0.500 & -1.516 & -2.516 \\
    $c_0$      &  1.006 & 1.004 & 1.003 &  1.005 &  1.003  & 1.001 &  1.004 & 1.002  & 1.001
  \end{tabular}
  \caption{
    Growth rate, frequency, wavenumber and phase velocity for the most unstable modes of a
    uniform helix for $H/R=\pi/5$ and different core sizes. Here, $\Delta k =
    0.01625$.
  }
  \label{newtab:1}
\end{table}
%%%%%%%%%%%%%%%%%%%%%%%%%%%%%%%%%%%%%%%%%%%%%%%%%%%%%%%%%%%%%%%%%%%%%%%%%%%%%%%%
%%%%%%%%%%%%%%%%%%%%%%%%%%%%%%%%%%%%%%%%%%%%%%%%%%%%%%%%%%%%%%%%%%%%%%%%%%%%%%%%

%%%%%%%%%%%%%%%%%%%%%%%%%%%%%%%%%%%%%%%%%%%%%%%%%%%%%%%%%%%%%%%%%%%%%%%%%%%%%%%%
%%%%%%%%%%%%%%%%%%%%%%%%%%%%%%%%%%%%%%%%%%%%%%%%%%%%%%%%%%%%%%%%%%%%%%%%%%%%%%%%

\begin{figure}\it\centering
  \includegraphics[scale=0.9, trim=2 2 2 2, clip]{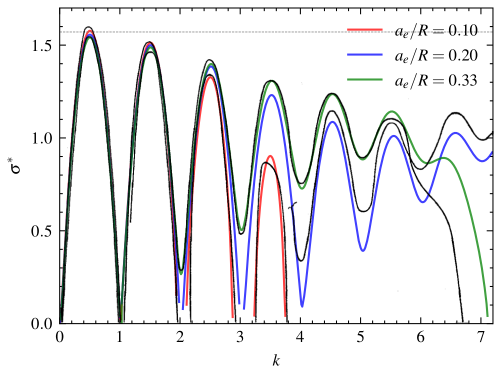}
  \caption{
    Dimensionless growth rate $\sigma^*$ of a uniform helix as function of $k_0$
    for $H/R=\pi/5$ and different core sizes. Black thin lines correspond to
    figure 5d reproduced from \cite{widnall1972stability}, while colour lines
    correspond to our numerical approach.
  }
  \label{newfig:4}
\end{figure}
%%%%%%%%%%%%%%%%%%%%%%%%%%%%%%%%%%%%%%%%%%%%%%%%%%%%%%%%%%%%%%%%%%%%%%%%%%%%%%%%
%%%%%%%%%%%%%%%%%%%%%%%%%%%%%%%%%%%%%%%%%%%%%%%%%%%%%%%%%%%%%%%%%%%%%%%%%%%%%%%%

The (temporal) frequency spectra for a helix of pitch $H/R=\pi/5$ and
$a_e/R=0.1$ is presented in figure \ref{newfig:3}a. By construction, the
spectrum is symmetric with respect to zero, i.e.
$\sigma^*(\omega^*)=\sigma^*(-\omega^*)$, and only the positive frequencies are
displayed. The most unstable frequencies are located near $\omega^*=3$ with
additional local maxima at odd multiples of this frequency (see left panel in
table \ref{newtab:1}). Figure \ref{newfig:3}b displays the Fourier spectrum of
$\check{z}_m$ for the most unstable mode, which is characterized by a single
peak at $k=-0.500$. We can show that each eigenvector has the form of a complex
wave with a single dominant wavenumber $k=k_0$, such that a direct
correspondence between $\sigma_m$, $\omega_m$ and $k$ can be established, see
figure \ref{newfig:3}c.

Axial perturbations propagate along the structure as the sum of travelling waves
\begin{equation}\label{eq:0.2}
  z'(\zeta,\psi) = \sum_m \hat{z}_m \cos(\omega_m \psi + k_0\zeta/C_\theta) e^{\sigma_m \psi}
\end{equation}
with phase velocity
\begin{equation}
  c_0 = -\omega_m C_\theta/k_0.
\end{equation}
A similar behaviour is observed for radial and azimuthal perturbations. As noted
by \cite{brynjell2020numerical}, the frequencies obtained in a frame rotating
with $\Omega_R$ can be mapped into a second reference frame rotating with
$\Omega_R'$ through
\begin{equation}
  \omega_m' = \omega_m + (\Omega_R'-\Omega_R) k/C_\theta.
\end{equation}

For instance, in our example most of the tangential velocity comes from the
moving frame, i.e., $ \vert\Omega\vert \ll \vert\Omega - \Omega_R\vert$, such
that perturbations are advected with $c_0$ close to 1, see table \ref{newtab:1}.
If, instead, we consider a frame moving with the vortex elements, perturbations
are advected with $c_0$ close to 0 for the same wavenumber (figure
\ref{newfig:3}d).

{%
As seen in figure \ref{newfig:4}, our numerical results are in good agreement
with the stability curves presented in figure 5d by \cite{widnall1972stability}
and figure 3 by \cite{quaranta2015long} for the same parameters. As mentioned in
section \ref{subsec:linear}, our approach differs from previous works in the way
the Jacobian matrix is evaluated. As long as the base flow is (spatially)
periodic, it is allowed to take any shape since equation \eqref{sec:eqlinear} is
evaluated directly from the discretised vortex segments. This will be useful
for studying the more geometrically challenging helical pairs.
}

\section{Stability of one vortex pair without central hub vortex}
\label{sec4}

{
In this section, we describe the unstable modes for the case of one vortex pair
without central hub vortex. Depending on the geometric parameters, some modes
become more prominent than others. For clarity, we introduce progressively the
unstable modes for {\it (i)} leapfrogging, {\it (ii)} sparsely braided and {\it
(iii)} densely braided wakes. Leapfrogging wakes display two types of unstable
modes: the pairing of the large-scale pattern and a new type specific to this
configuration. Sparsely braided wakes display an additional type, which becomes
more prominent as $\beta$ increases. Finally, densely braided wakes display an
additional type, which corresponds to the pairing modes of the vortex pair.
}

\subsection{Typical displacement modes for leapfrogging wakes}
\label{sec4.1}
For each pair, we introduce the following decomposition
\begin{equation}\label{eq:2.1}
  \boldsymbol{X}_+ \equiv (\boldsymbol{X}_1+\boldsymbol{X}_2)/2, \quad\quad
  \boldsymbol{X}_- \equiv (\boldsymbol{X}_1-\boldsymbol{X}_2)/2
\end{equation}
where $\boldsymbol{X}_+$ characterises the large-scale pattern traced by the
vorticity barycentre, while $\boldsymbol{X}_-$ represents the rotation of the
vortex pair relative to $\boldsymbol{X}_+$. Figure \ref{newfig:1}b depicts a
leapfrogging wake with $\beta=p/q=3/4$, where $\boldsymbol{X}_+$ is represented
as a tube enclosing the vortex pair. {Over a single period, the vortex pair
completes $p=3$ rotations around $\mathscr{H}$, while $\mathscr{H}$ completes
$q=4$ rotations around the $z$ axis.} Note that $\boldsymbol{X}_+ $ and
$\mathscr{H}$ are close but not exactly equal due to the effect of
self-induction. In a similar vein, we introduce the following decomposition
\begin{equation}
  \boldsymbol\phi_m^{+} \equiv (\boldsymbol\phi_m^{(1)} + \boldsymbol\phi_m^{(2)})/2, \quad\quad
  \boldsymbol\phi_m^{-} \equiv (\boldsymbol\phi_m^{(1)} - \boldsymbol\phi_m^{(2)})/2
\end{equation}
where $\boldsymbol\phi_m^{+}$ and $\boldsymbol\phi_m^{-}$ indicate the
displacement modes of $\boldsymbol{X}_+$ and $\boldsymbol{X}_-$, respectively.

\begin{figure}\it
  \begin{minipage}[c]{0.85\linewidth}\it
    \hspace{0.5cm} (a) \hfill (b) \hfill~ \\
    \includegraphics[width=\linewidth, trim=0 0 0 0, clip]{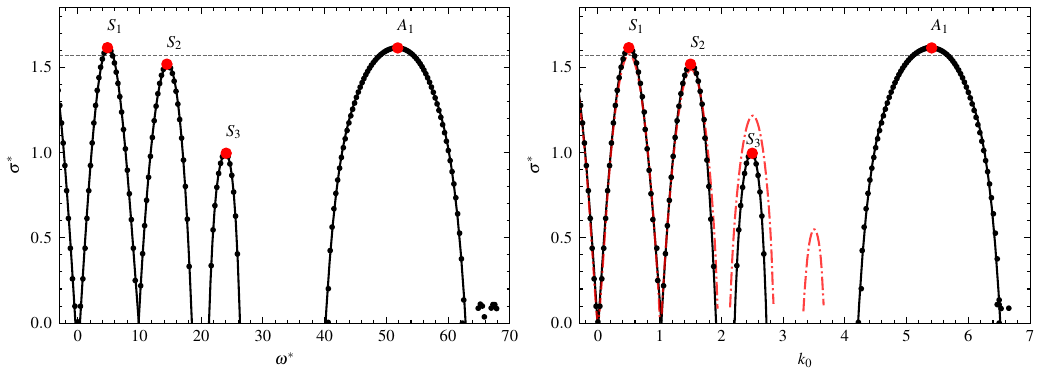}
  \end{minipage}
  \begin{minipage}[c]{0.14\linewidth}
    (c)

    \centering
    S-Modes \\

    \includegraphics[scale=0.30, trim= 50 60 265 -10, clip]{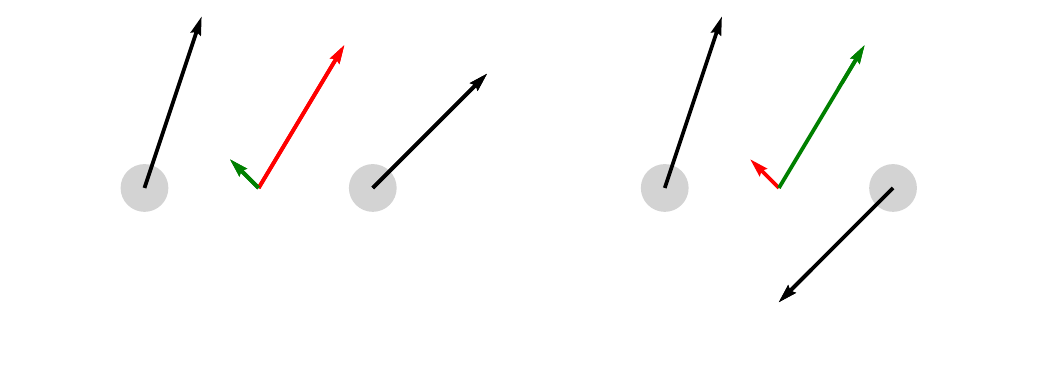} \\

    A-Modes \\
    \includegraphics[scale=0.30, trim=305 0 20  -10, clip]{./new_figures/figure6c.pdf} \\
  \end{minipage}

  \caption{
    Stability curves of a leapfrogging wake ($R^*=7$, $H^*=5.25$, $\beta=3/4$):
    Dimensionless growth rate $\sigma^*$ as function of \emph{(a)} $\omega^*$
    and \emph{(b)} $k_0$. For reference, fig. \emph{(b)} shows the prediction
    for a uniform helix ($H/R=0.75$, $a_e/R=0.1$) in red lines. A schematic
    representation of the two groups of modes is shown on fig. \emph{(c)},
    where the symmetric (resp. anti-symmetric) part is shown in red
    (resp. green) arrows.
   }
  \label{newfig:5}
\end{figure}

Figure \ref{newfig:5} presents the frequency spectrum, which is characterized by
a set of modes distributed over three contiguous lobes at low frequencies, and a
second set of modes over an additional lobe at higher frequencies. The maximum
growth rate is observed near $\omega^*=4.8$, with additional local maxima (in
descending order) near $\omega^*=51.9$, $\omega^*=14.5$, and $\omega^*=24.0$.
These values respectively correspond to dominant wavenumbers $k_0=0.5$,
$k_0=5.4$, $k_0=1.5$, and $k_0=2.5$.  Dimensionless growth rates are slightly
larger than the equivalent helical vortex (figure \ref{newfig:5}b). This can be
explained by the change in the effective distance separating neighbouring loops.

\begin{figure}\it
  (a) \hfill (b) \hfill ~ \\
  \includegraphics[width=0.495\linewidth, trim=0 0 -10 0, clip]{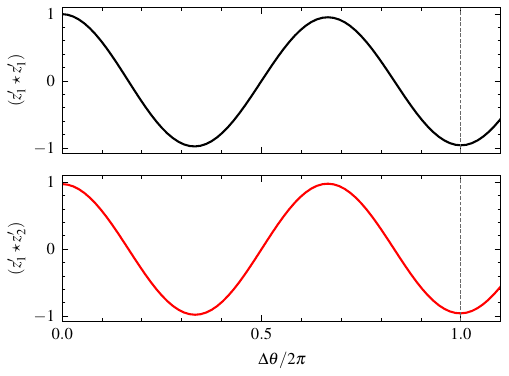}
  \includegraphics[width=0.495\linewidth, trim=0 0 -10 0, clip]{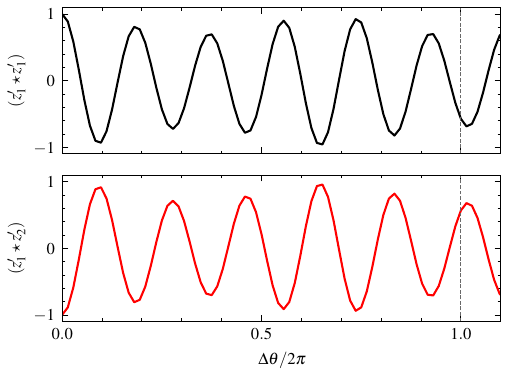}
  \caption{
    Function $(z_1'\star z_1')$ (top) and $(z_1'\star z_2')$ (bottom) for the
    case in figure \ref{newfig:1}b and modes: \emph{(a)} $S_2$, and \emph{(b)}
    $A_1$.
  }
  \label{newfig:10}
\end{figure}

{%
Low frequency modes are clearly reminiscent of the unstable modes for the
equivalent uniform helices, while those at higher frequencies are specific to
this geometry. The two groups differ in the relative alignment between the
displacements of the pair: predominantly aligned displacements (or symmetric
with respect to the vorticity barycentre, figure \ref{newfig:5}c top) for low
frequency modes and predominantly opposed displacements (or anti-symmetric with
respect to the vorticity barycentre, figure \ref{newfig:5}c bottom) for those at
higher frequencies.
A more quantitative way to illustrate this difference is through the spatial
cross-correlation
\begin{equation}
  (z_i' \star z_j')(\Delta\theta) \equiv \int_\infty^\infty z_i'(\zeta/C_\theta) z_j'(\zeta/C_\theta + \Delta\theta) d\zeta
\end{equation}
where $\Delta\theta$ is the delay in angular position. For instance, for $S_2$
the displacements between neighbouring turns, i.e. $\Delta\theta=2\pi$, are well
anti-correlated since perturbations are in opposition of phase (figure
\ref{newfig:10}a). Conversely, for mode $A_1$ the auto-correlation between
consecutive turns is negative, while the cross-correlation is positive (figure
\ref{newfig:10}b), meaning that vortices move in opposite directions, but one of
them is aligned with one of the vortices in the neighbouring loop.
}

\begin{figure}\centering\it
  \hspace{1.0cm} (a) \hfill (b) \hfill ~\\
  \includegraphics[width=0.9\linewidth, trim=0 0 0 0, clip]{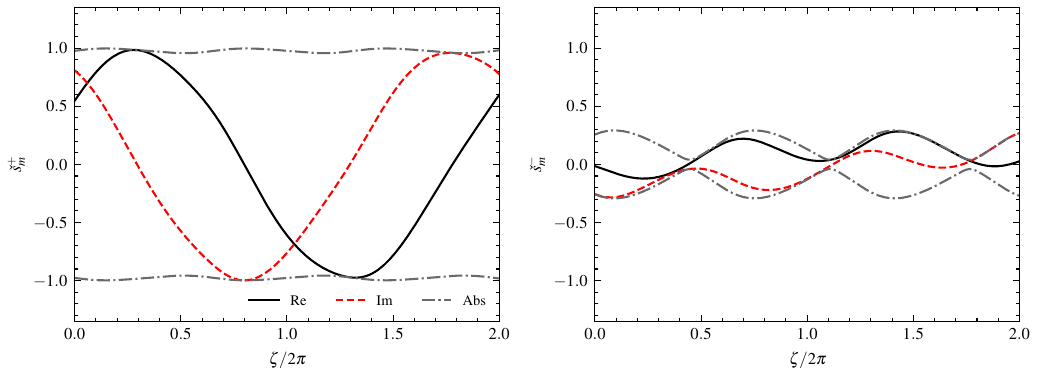}\\

  \hspace{1.0cm} (c) \hfill (d) \hfill ~\\
  \includegraphics[width=0.9\linewidth, trim=0 0 0 0, clip]{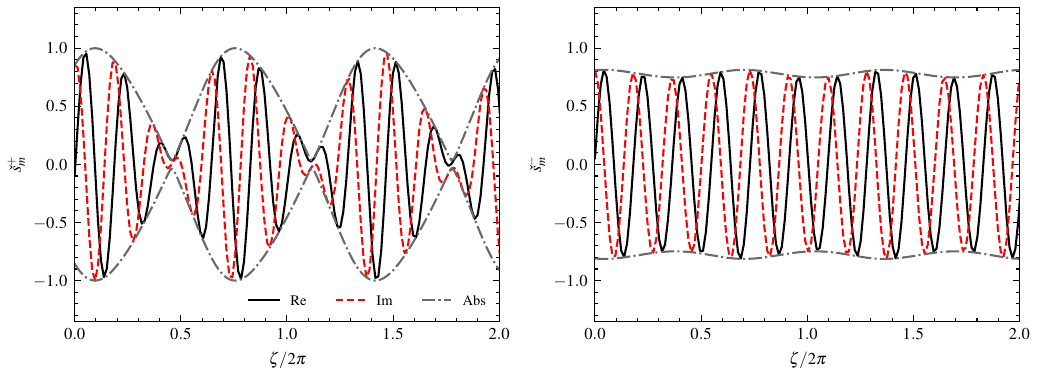}\\
  \caption{
    Complex eigenvectors for the case in figure \ref{newfig:5}. Perturbations in
    the local axial direction \emph{(a,c)} $\check{s}_{m}^{+}$, and \emph{(b,d)}
    $\check{s}_{m}^{-}$ shown for modes \emph{(a,b)} $S_1$, and \emph{(c,d)}
    $A_1$.
  }
  \label{newfig:6}
\end{figure}

Figure \ref{newfig:6} shows the eigenvectors
$\check{s}_{m}^{+}$ and $\check{s}_{m}^{-}$ corresponding to the most unstable
modes $S_1$ and $A_1$. {Here, $\check{s}_{m}^{+}$ (resp. $\check{s}_{m}^{-}$ )
corresponds to the component of $\boldsymbol\phi_m^{+}$ (resp.
$\boldsymbol\phi_m^{-}$) along the local axial direction.} For the symmetric
mode, $\check{s}_{m}^{+}$ is dominant and has a nearly constant envelope, while
$\check{s}_{m}^{-}$ has a sinusoidal envelope, whereas the anti-symmetric mode
displays the opposite behaviour.

%%%%%%%%%%%%%%%%%%%%%%%%%%%%%%%%%%%%%%%%%%%%%%%%%%%%%%%%%%%%%%%%%%%%%%%%%%%%%%%%
%%%%%%%%%%%%%%%%%%%%%%%%%%%%%%%%%%%%%%%%%%%%%%%%%%%%%%%%%%%%%%%%%%%%%%%%%%%%%%%%
%%%%%%%%%%%%%%%%%%%%%%%%%%%%%%%%%%%%%%%%%%%%%%%%%%%%%%%%%%%%%%%%%%%%%%%%%%%%%%%%
%%%%%%%%%%%%%%%%%%%%%%%%%%%%%%%%%%%%%%%%%%%%%%%%%%%%%%%%%%%%%%%%%%%%%%%%%%%%%%%%
%%%%%%%%%%%%%%%%%%%%%%%%%%%%%%%%%%%%%%%%%%%%%%%%%%%%%%%%%%%%%%%%%%%%%%%%%%%%%%%%
%%%%%%%%%%%%%%%%%%%%%%%%%%%%%%%%%%%%%%%%%%%%%%%%%%%%%%%%%%%%%%%%%%%%%%%%%%%%%%%%
%%%%%%%%%%%%%%%%%%%%%%%%%%%%%%%%%%%%%%%%%%%%%%%%%%%%%%%%%%%%%%%%%%%%%%%%%%%%%%%%
%%%%%%%%%%%%%%%%%%%%%%%%%%%%%%%%%%%%%%%%%%%%%%%%%%%%%%%%%%%%%%%%%%%%%%%%%%%%%%%%
%%%%%%%%%%%%%%%%%%%%%%%%%%%%%%%%%%%%%%%%%%%%%%%%%%%%%%%%%%%%%%%%%%%%%%%%%%%%%%%%
\begin{figure}\it
  \hspace{0.5cm} (a) \hfill (b) \hfill (c) \hfill ~ \\
  \includegraphics[width=\linewidth, trim=0 0 0 11, clip]{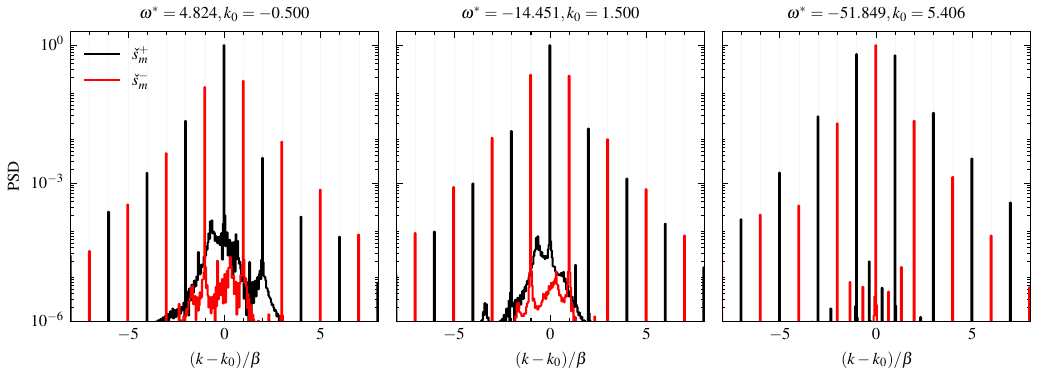}

  \caption{
    Fourier spectrum of modes: \emph{(a)} $S_1$, \emph{(b)} $S_2$, and \emph{(c)} $A_1$,
    for the case in figure \ref{newfig:5}.
  }
  \label{newfig:7}
\end{figure}
%%%%%%%%%%%%%%%%%%%%%%%%%%%%%%%%%%%%%%%%%%%%%%%%%%%%%%%%%%%%%%%%%%%%%%%%%%%%%%%%
%%%%%%%%%%%%%%%%%%%%%%%%%%%%%%%%%%%%%%%%%%%%%%%%%%%%%%%%%%%%%%%%%%%%%%%%%%%%%%%%
%%%%%%%%%%%%%%%%%%%%%%%%%%%%%%%%%%%%%%%%%%%%%%%%%%%%%%%%%%%%%%%%%%%%%%%%%%%%%%%%
\begin{table}\centering
  \begin{tabular}{crrcccr}
    Mode  & Frequency & \multicolumn{4}{c}{Wavenumbers} & Phase velocity\\
          & $\omega^*$& $k_0$    & $k_1$             & $k_2$             & $k_3$             & $c_0$   \\
    $S_1$ & $ 4.824$   & $-0.500$ & $(-1.250, +0.250)$ & $(-2.000,+1.000)$ & $(-2.750,+1.750)$ & $1.002$ \\
    $S_2$ & $14.451$   & $-1.500$ & $(-2.250, -0.750)$ & $(-3.000,+0.000)$ & $(-3.750,+0.750)$ & $1.000$ \\
    $A_1$ & $51.849$   & $-5.406$ & $(-6.156, -4.656)$ & $(-6.906,-3.906)$ & $(-7.656,-3.156)$ & $0.996$
  \end{tabular}
  \caption{Leading wavenumbers of the modes shown in figure \ref{newfig:7}. Here, $\Delta k = 0.03125$.}
  \label{newtab:2}
\end{table}
%%%%%%%%%%%%%%%%%%%%%%%%%%%%%%%%%%%%%%%%%%%%%%%%%%%%%%%%%%%%%%%%%%%%%%%%%%%%%%%%
%%%%%%%%%%%%%%%%%%%%%%%%%%%%%%%%%%%%%%%%%%%%%%%%%%%%%%%%%%%%%%%%%%%%%%%%%%%%%%%%

Each mode displays a dominant wavenumber $k_0$ with additional peaks (typically in
descending order of magnitude) at wavenumbers $k_n \approx k_0 \pm n \beta$ for
integer values of $n$ (see Fourier spectra in figures \ref{newfig:7}a-c and
table \ref{newtab:2}). This pattern is also observed in the spectra of the axial
and radial components, $\check{z}_{m}$ and $\check{r}_{m}$. From these observations, we infer that
perturbations propagate along the structure as
\begin{equation}\label{eq:2.2}
  s'_j(\zeta,\psi) \approx \sum_m \left[\bigg[ \hat{s}_{m0}  + \sum_{n=1} \hat{s}_{mn} \cos{(n\beta\zeta/C_\theta + \hat{\varphi}_{mn})} \bigg] \cos(\omega_m \psi + k_0\zeta/C_\theta) e^{\sigma_m \psi} \right]
\end{equation}
where $\hat{s}_{mn}$ and $\hat{\varphi}_{mn}$ are the amplitudes and phase
differences measured with respect to $k_0$. Consider the leading terms in
\eqref{eq:2.2}. For symmetric modes $s_+^\prime$ roughly corresponds to a
travelling wave, whereas $s_-^\prime$ approximates a wave that is modulated in
amplitude by a cosine function of period $\zeta_B$, i.e., the periodicity of the
base flow (figures \ref{newfig:7}a-b). Anti-symmetric modes display the opposite
behaviour: $s_-^\prime$ approximates a travelling wave, whereas $s_+^\prime$ is
modulated in amplitude by a function of period $\zeta_B$ (figure
\ref{newfig:7}c). In both cases, the contributions from higher order terms also
correspond to waves modulated in amplitude by multiples of $\zeta_B$ in
decreasing order of magnitude.

%%%%%%%%%%%%%%%%%%%%%%%%%%%%%%%%%%%%%%%%%%%%%%%%%%%%%%%%%%%%%%%%%%%%%%%%%%%%%%%%
\begin{figure}
  \includegraphics[width=\linewidth, trim=0 0 0 0, clip]{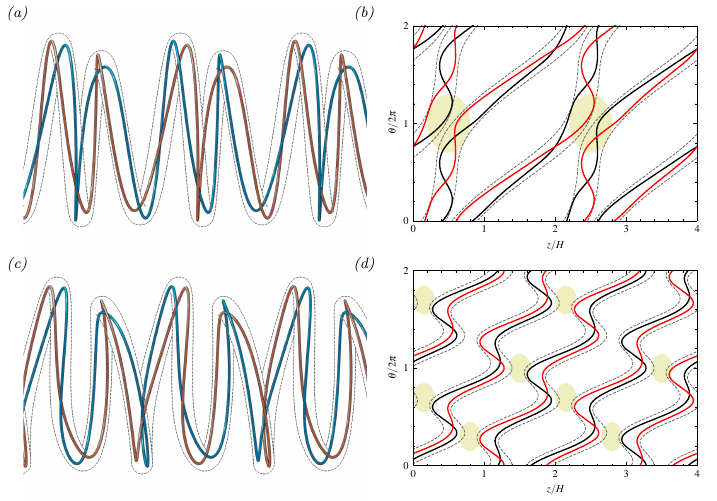}
  \caption{
    Base state from figure \ref{newfig:1}b perturbed by modes
    \emph{(a, b)} $S_1$ and \emph{(c, d)} $S_2$. \emph{(a, c)} Three dimensional
    and \emph{(b, d)} developed plan views illustrating the local pairing modes.
  }
  \label{newfig:8}
\end{figure}
%%%%%%%%%%%%%%%%%%%%%%%%%%%%%%%%%%%%%%%%%%%%%%%%%%%%%%%%%%%%%%%%%%%%%%%%%%%%%%%%
%%%%%%%%%%%%%%%%%%%%%%%%%%%%%%%%%%%%%%%%%%%%%%%%%%%%%%%%%%%%%%%%%%%%%%%%%%%%%%%%
\begin{figure}\centering\it
  \includegraphics[width=\linewidth, trim=0 0 0 0, clip]{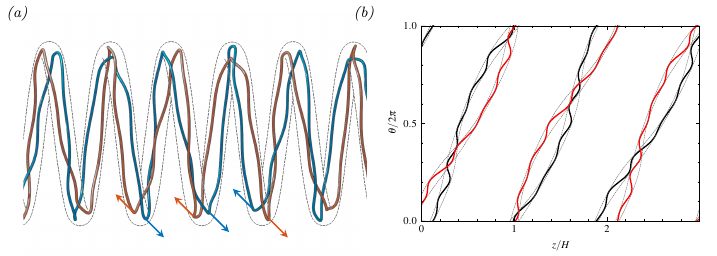}
  \caption{
  Base state from figure \ref{newfig:1}b perturbed by mode $A_1$. \emph{(a)}
  Three dimensional and \emph{(b)} developed plan view. Arrows indicate the
  displacements at the center plane in fig. \emph{(a)}.
  }
  \label{newfig:9}
\end{figure}
%%%%%%%%%%%%%%%%%%%%%%%%%%%%%%%%%%%%%%%%%%%%%%%%%%%%%%%%%%%%%%%%%%%%%%%%%%%%%%%%

Figure \ref{newfig:8}a presents the deformation due to the symmetric mode $S_1$
by plotting the perturbed geometry $\boldsymbol{X}_j=\boldsymbol{X}_j^B +
\boldsymbol{X}_j'$ for some arbitrary amplitude. A developed plan view
illustrates $p=1$ localized pairing events for every $q=2$ neighbouring turns of
the large-scale pattern (seen as dashed lines in figures \ref{newfig:8}a-b). An
additional example corresponding to mode $S_2$ is shown in figures
\ref{newfig:8}c-d, where $p=3$ localized pairing events are observed every $q=2$
neighbouring turns. This behaviour was expected since the predominantly aligned
displacements result in a block displacement of the vortex pair. As a result,
the large-scale pattern behaves like a uniform helix where perturbations with
wavenumber $k_0=p/q$ repeat after $p$ cycles and display local pairing events at
$q$ azimuthal locations (\cite{widnall1972stability}).

Anti-symmetric modes behave in a different manner. Here, the two vortices move
towards (or away from) one another such that $\boldsymbol{X}_+$ deforms much
less and only at specific positions (figure \ref{newfig:9}a). In other words,
the pair predominantly displays an anti-symmetric motion with respect to the
helical structure, hence the name. Displacements are localized and not
necessarily aligned with the rotation of the pair. For instance, at the azimuths
where displacements are perpendicular to the line connecting the two vortices
the structure is twisted back and forth, whereas when displacements are
parallel, the separation distance expands and contracts. The latter could
potentially trigger the merging of the vortex pair. This localization can be
deduced from the envelopes of the corresponding eigenvectors, illustrated as
dashed lines enclosing each vortex in figure \ref{newfig:9}b. If we consider a
longitudinal cut, we can see that displacements of a given vortex are paired
with one of the vortices in the neighbouring turn but not with its companion
(see arrows at the bottom part of figure \ref{newfig:9}a). However, the choice
of the characteristic wavenumber $k_0$ is not obvious. For instance, doubling
$H^*$ from $5.25$ to $10.5$ shifts mode $A_1$ from $k_0=5.406$ to $k_0=5.938$,
while increasing $\beta$ from $3/4$ to $7/2$ shifts the wavenumber to
$k_0=5.552$. However, perturbations between consecutive pairs remain
anti-correlated, suggesting the dominant wavenumber is selected by the geometry
as the one that amplifies the local pairing. We shall explore this relation in
the following section.

In the appendix \ref{secA}, we also analyse the long-wave instability of our
solutions using the linear impulse response approach developed in
\cite{venegas2020}. The idea is to solve the linear perturbation equation
\eqref{sec2:eq16} using a Dirac as initial condition and analyse the
spatio-temporal growth of the resulting wave packet. A sufficiently long domain
is considered so that the wavepacket does not reach the boundaries during the
length of the simulations. As shown in the appendix, the
temporal modes can be recovered but their study is more difficult to perform
with the linear impulse response as all the instability modes are simultaneously
excited. However, the linear impulse response provides additional information
by telling us how the instability spreads in space. In particular, we are able to
show that although the most unstable anti-symmetric perturbations propagate at a
similar speed than the symmetric perturbations, their spreading in space is much
less important. The transition from convective to absolute instability is
therefore expected to be associated with the symmetric perturbations. Moreover,
as for the linear spectrum, we also demonstrate that the spatio-temporal
evolution of these symmetric perturbations can be well-described by the
spatio-temporal evolution of the linear impulse response on a uniform helical
vortex of large core size.

\subsection{Influence of $H^*$ and $R^*$ {on the modes of leapfrogging wakes}}

%%%%%%%%%%%%%%%%%%%%%%%%%%%%%%%%%%%%%%%%%%%%%%%%%%%%%%%%%%%%%%%%%%%%%%%%%%%%%%%%
\begin{figure}\it
  \hspace{0.25cm} (a) \hfill (b) \hfill (c) \hfill ~ \\
  \includegraphics[width=\linewidth, trim=0 0 0 0, clip]{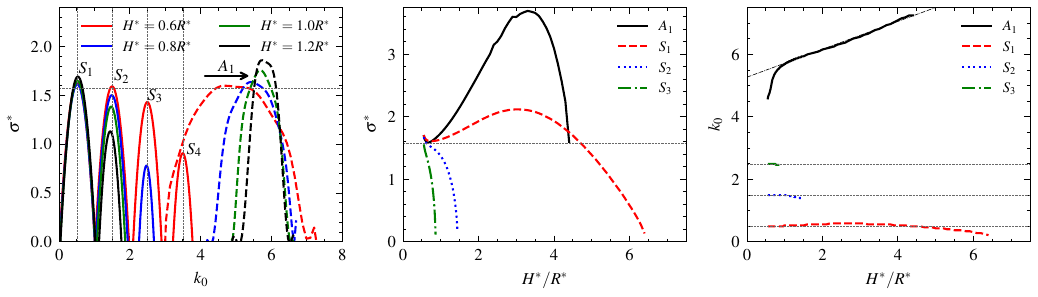}

  \hspace{0.25cm} (d) \hfill (e) \hfill (f) \hfill ~ \\
  \includegraphics[width=\linewidth, trim=0 0 0 0, clip]{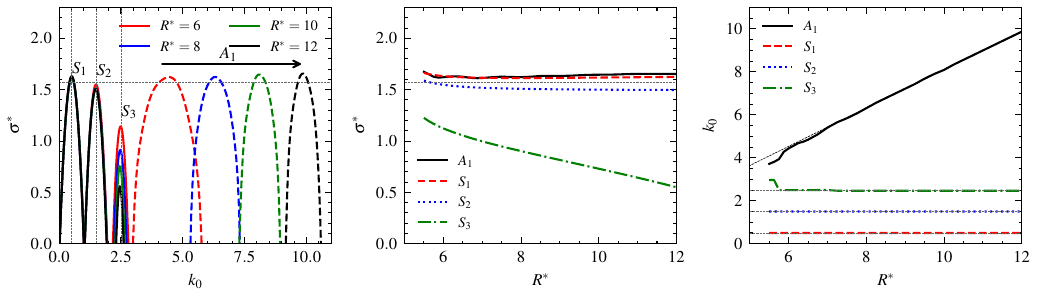}
  \caption{
    Growth rates and dominant wavenumbers for $\beta=3/4$. Figs.
    {\it (a,b,c)} shown as function of $H^*/R^*$ for $R^*=7$; and
    {\it (d,e,f)} shown as function of $R^*$ for $H^*/R^*=0.75$.
  }
  \label{newfig:11}
\end{figure}
%%%%%%%%%%%%%%%%%%%%%%%%%%%%%%%%%%%%%%%%%%%%%%%%%%%%%%%%%%%%%%%%%%%%%%%%%%%%%%%%

Since the two kinds of pairing modes described in section \ref{sec4.1} seem to
involve neighbouring turns of $\mathscr{H}$, growth rates are expected to
display a strong dependency on the relative pitch $H^*/R^*$. In general, the
maximum growth rate is larger than the maximum growth rate obtained for uniform
helices with the total circulation of the vortex pair (figure \ref{newfig:11}a),
which is slightly larger than $\sigma=(\pi/2)(\Gamma/H^2)$ obtained for an array
of point vortices of circulation $2\Gamma$ and separated by a distance $H$. For
the range of values considered, symmetric and anti-symmetric modes have
comparable growth rates (figure \ref{newfig:11}a-b). For symmetric modes,
$\sigma^*$ initially increases for mode $S_1$ before vanishing, while the modes
$S_2$, $S_3$, and so on, gradually vanish as the pitch increases. For
anti-symmetric modes, $\sigma^*$ initially increases at a faster rate, but also
vanishes more quickly, while the dominant wavenumber is approximated by
$k_0=0.45H^*/R^*$ for $H^*/R^*>2$, see figure \ref{newfig:11}c.

It is also interesting to fix the relative pitch $H^*/R^*$ and change the
separation distance through $R^*\equiv R/d$ (figure \ref{newfig:11}d).
Stability curves are essentially unchanged for modes $S_1$ and
$S_2$, while the growth rate of $S_3$ is observed to decrease as the separation distance
becomes smaller (figure \ref{newfig:11}c). This is also reminiscent of uniform
helical vortices, where the maximum growth rates also decrease as the effective
core size becomes smaller (see, for instance figure \ref{newfig:4}). For mode
$A_1$, the growth rate remains constant (figure \ref{newfig:11}e), while $k_0$
seems to evolve linearly with $k_0 \sim 0.89R^*$ and overlaps the symmetric
modes for $R^*<6$ (figure \ref{newfig:11}f).

{%
From this dependency on $H^*$, and $R^*$, we may deduce the following: {\it (i)}
$\sigma$ scales with $\Gamma/H^2$ for modes $A_1$ and $S_1$ whatever $H^*$ and
$R^*$. This is consistent with a local pairing acting over a distance
comparable to $H$; and {\it (ii)} the wavenumber most amplified by $A_1$
increases linearly with $H^*/R^*$ (for constant $R^*$) and with $R^*$ (for constant
$H^*/R^*$). In other words, mode $A_1$ deviates from the classical pairing of
uniform helices, and amplifies a linear wavelength $l_0=2\pi R k_0 \sim d^{-1}$
instead. This is reminiscent of a four-vortex system involving two co-rotating
pairs separated by a distance $b=O(H)$ (see, for instance
\cite{crouch1997instability, fabre2000stability}). We shall revisit this matter
in section \ref{sec5} with the case of $N=2$ helical pairs.
}

{%
\subsection{Typical displacement modes for sparsely braided wakes}

%%%%%%%%%%%%%%%%%%%%%%%%%%%%%%%%%%%%%%%%%%%%%%%%%%%%%%%%%%%%%%%%%%%%%%%%%%%%%%%%
\begin{figure}\it\centering
  \hspace{0.5cm} (a) \hfill\hfill\hspace{1.5cm} (b) \hspace{1.5cm}\hfill ~ \\
  \includegraphics[scale=0.75, trim=0 0 0 0, clip]{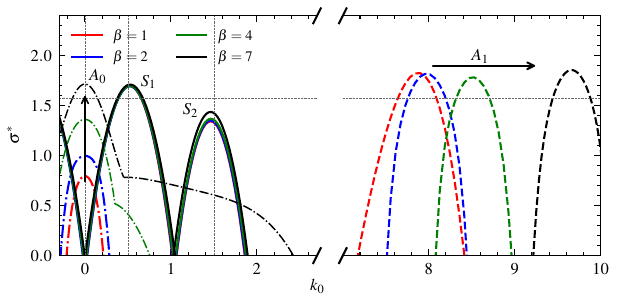}
  \includegraphics[scale=0.75, trim=0 0 0 0, clip]{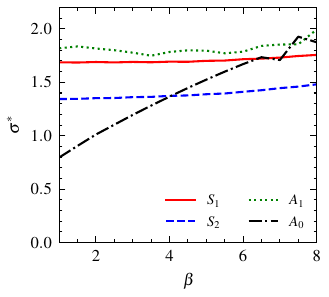}
  \caption{
    Stability of sparsely braided wakes for $R^*=10$ and $H^*/R^*=1$.
    {\it (a)} Growth rates and dominant wavenumbers;
    and {\it (b)} maximum growth rates as function of $\beta$.
  }
  \label{newfig:12}
\end{figure}
%%%%%%%%%%%%%%%%%%%%%%%%%%%%%%%%%%%%%%%%%%%%%%%%%%%%%%%%%%%%%%%%%%%%%%%%%%%%%%%%
%%%%%%%%%%%%%%%%%%%%%%%%%%%%%%%%%%%%%%%%%%%%%%%%%%%%%%%%%%%%%%%%%%%%%%%%%%%%%%%%
\begin{figure}\it\centering
  \hspace{1.0cm} (a) \hfill (b) \hspace{1.5cm}\hfill ~ \\
  \includegraphics[scale=0.75, trim=0 2 250 378, clip]{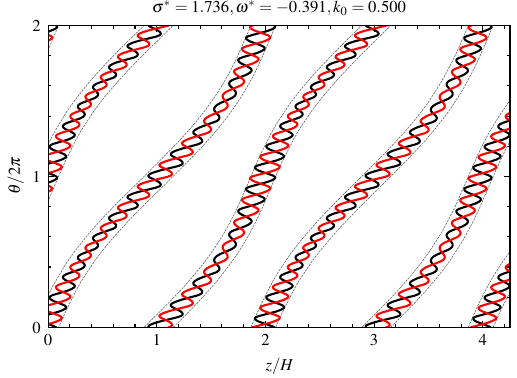}
  \includegraphics[scale=0.75, trim=0 380 250 0, clip]{./new_figures/figure13.pdf}
  \caption{
    Base state from figure \ref{newfig:12} and $\beta=7$
    perturbed by modes {\it(a)} $S_1$ and {\it(b)} $A_0$. Figure
    {\it(b)} inset includes the base state (in dashed lines) for reference.
  }
  \label{newfig:13}
\end{figure}
%%%%%%%%%%%%%%%%%%%%%%%%%%%%%%%%%%%%%%%%%%%%%%%%%%%%%%%%%%%%%%%%%%%%%%%%%%%%%%%%

Figure \ref{newfig:12}a presents the stability curves as we move from
leapfrogging to sparsely braided wakes ($1<\beta<10$). Modes $S_1$, $S_2$,
and $A_1$, behave as in leapfrogging wakes. Some modes may act on the
large-scale pattern and on the distance separating the vortex pair (see, for
instance mode $S_1$ in figure \ref{newfig:13}a). For these modes, small
oscillations in $\sigma^*(\beta)$ can be explained by a change in the effective
distance between neighbouring vortices, which varies by a factor $d\cos(\phi)$
due to relative orientation of the vortex pairs, where $\phi = 2\pi/q$ for
$\beta=p/q$. For $S_1$ and $S_2$, the unstable frequencies and leading
wavenumbers are unchanged, while the wavenumber most amplified by $A_1$ displays
some dependency on $\beta$.

We observe an additional set of low frequency modes (figure \ref{newfig:12}a).
Of this set of modes, the maximum growth corresponds to the zero-frequency mode
$A_0$, which is almost a linear function of $\beta$ (figure \ref{newfig:12}b).
The displacement produced by $A_0$ is characterized by a radial expansion of the
vortex pair and a translation along the helical coordinate (see, developed plan
view in figure \ref{newfig:13}b and corresponding inset). The resulting perturbed state would correspond
to a similar braided wake but one with slightly larger $d$ and $R$. Other modes
in the same branch display a similar displacement, but one where radial
expansion and the translation along $\mathscr{H}$ are modulated in amplitude by
a wavenumber $k_0$. For the case of two uniform helices, an equivalent
displacement would yield two uniform helices of slightly larger radius. In such
a case, the initial displacement is not amplified and the system remains in
neutral equilibrium. However, for helical braids, the perturbed state is not
necessarily a solution of \eqref{sec2:eq1} explaining the positive growth rate.
This branch is not always present and, given the size of the parameter space
($H^*$, $R^*$, and $\beta$), it is unclear how $\sigma^*$ varies. For instance,
for the case presented in figure \ref{newfig:12}b and $\beta=4$,
$\sigma^*=\sigma (H^2/\Gamma)$ is close to $1.5$. For the same $(R^*, \beta)$
and $H^*/R^*=2$, $\sigma^*$ is nearly four times larger, suggesting these
modes no longer scale with $\Gamma/H^2$. This behaviour extends to the case of
densely braided wakes.

\subsection{Typical displacement modes for densely braided wakes}

%%%%%%%%%%%%%%%%%%%%%%%%%%%%%%%%%%%%%%%%%%%%%%%%%%%%%%%%%%%%%%%%%%%%%%%%%%%%%%%%
\begin{figure}\it
  \includegraphics[width=\linewidth, trim=0 0 0 0, clip]{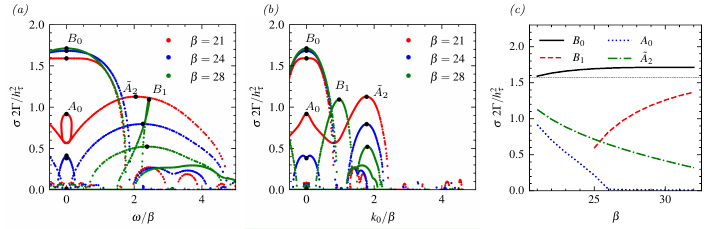}

  \caption{
    Stability of densely braided wakes for $R^*=10$ and $H^*/R^*=1$: growth rate
    as function of {\it(a)} $\omega$ and {\it(b)} $k_0$ for different $\beta$;
    and {\it(c)} growth rate of modes identified in figs. {\it(a,b)} as function
    of $\beta$.
  }
  \label{newfig:14}
\end{figure}
%%%%%%%%%%%%%%%%%%%%%%%%%%%%%%%%%%%%%%%%%%%%%%%%%%%%%%%%%%%%%%%%%%%%%%%%%%%%%%%%

%%%%%%%%%%%%%%%%%%%%%%%%%%%%%%%%%%%%%%%%%%%%%%%%%%%%%%%%%%%%%%%%%%%%%%%%%%%%%%%%
\begin{figure}\it\centering
  \includegraphics[width=0.95\linewidth, trim=0 0 0 0, clip]{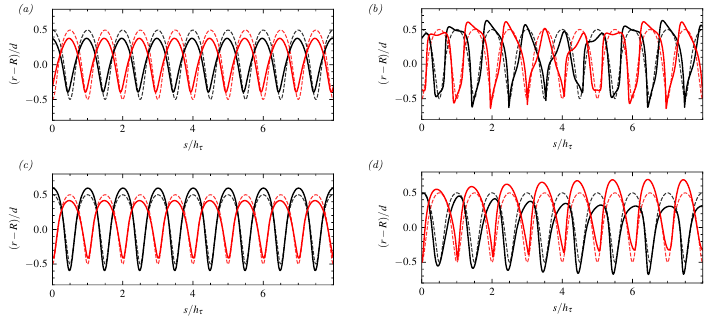}
  \caption{
    Stability of densely braided wakes. Schematic representation of the
    displacement modes identified in figure \ref{newfig:14}: {\it(a)} $A_0$,
    {\it(b)} $\tilde{A}_2$, {\it(c)} $B_0$, and {\it(d)} $B_1$.
    For reference, each figure also includes the base state (in dashed lines).
  }
  \label{newfig:15}
\end{figure}
%%%%%%%%%%%%%%%%%%%%%%%%%%%%%%%%%%%%%%%%%%%%%%%%%%%%%%%%%%%%%%%%%%%%%%%%%%%%%%%%

Densely braided wakes correspond to the case $\beta > 10$, when the pitch
$h_\tau$ becomes of the same order as the separation distance $d$ and the
pairing modes of the vortex pair become dominant. The stability curves in figure
\ref{newfig:14}a-b, display nearly all of the modes introduced so far. Modes
$S_1$, $S_2$, and $A_1$ remain unchanged but are dwarfed by the other modes. For
$\beta=21$, the branch containing mode $A_0$ is still present, and now contains
a new local maxima around $k_0/\beta=2$, denoted $\tilde{A}_2$ in figure
\ref{newfig:14}b. For $\beta=24$, the branch containing both modes splits in
two. Modes $A_0$ and $\tilde{A}_2$ display a similar scaling and seem to vanish
for large $\beta$ (figure \ref{newfig:14}c). The spatial structure of $A_0$ is
unchanged (see, figure \ref{newfig:13}b inset and \ref{newfig:15}a), while mode
$\tilde{A}_2$ displays a radial expansion modulated in amplitude with some
spatial frequency (figure \ref{newfig:15}b).

Finally, we have the pairing modes of the vortex pair. For these modes, the
scaling of the growth rate is different since the pairing acts over a distance
comparable to $h_\tau/2$ instead of $H$. The maximum growth rate is comparable to the
predictions for two interlaced helices and $\sigma = (\pi/2) (2\Gamma/h_\tau^2)$
obtained for an array of point vortices of circulation $\Gamma$ and separated by
$h_\tau/2$ (figure \ref{newfig:14}c). Here, mode $B_0$ corresponds to a special
case with $\omega=0$ and $k_0=0$. As shown in figure \ref{newfig:15}c,
displacements are characterized by a radial expansion of the vortex pair and a
translation along the helical coordinate for one vortex, and a radial
contraction and a translation in the opposite direction for the other one. This
results in a form of uniform pairing along the helical coordinate, analogous to
the global pairing mode of two uniform helices. For $\beta>25$, a second
maximum, denoted $B_1$, is observed near $k_0/\beta = 1$. As shown in figure
\ref{newfig:15}d, displacements approach the vortices in neighbouring turns at
specific intervals, analogous to the local pairing mode of two uniform helices.
}
% From this point of view, the most noticeable change concerns
% the phase velocity $c_0$ which decreases for large $\beta$ (figure
% \ref{newfig:12}c). For leapfrogging wakes $c_0$ is close to 1 but decreases as
% the structure becomes braided, indicating that vortex elements move with an
% angular velocity smaller than $\Omega_R$.

%\clearpage

\section{Stability of two vortex pair with central hub vortex}
\label{sec5}

A similar stability analysis can be performed for the case of two interlaced
vortex pairs with a central hub vortex, illustrated in figure \ref{newfig:1}c.
{
For this analysis, we considered the hub as a straight vortex. Contribution from
the hub are taken into account for the stability analysis but the hub itself was
not allowed to deform. We present only the modes corresponding to leapfrogging
wakes and focus on the differences with respect to the case of one helical pair.
Unstable modes associated with braided wakes are not discussed, but we expect them
to display roughly the same behaviour described in section \ref{sec4}.
}

\subsection{Typical displacement modes for leapfrogging wakes}

%%%%%%%%%%%%%%%%%%%%%%%%%%%%%%%%%%%%%%%%%%%%%%%%%%%%%%%%%%%%%%%%%%%%%%%%%%%%%%%%
\begin{figure}\it\centering

  \hspace{1.0cm} (a) \hfill (b) \hspace{1.5cm}\hfill ~ \\

  \includegraphics[scale=0.73, trim=250 0 0 0, clip]{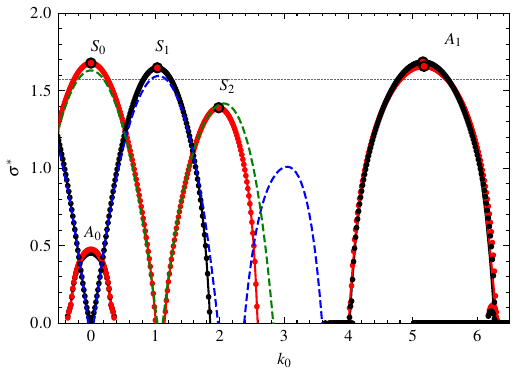}~
  \includegraphics[scale=0.73, trim=250 0 0 0, clip]{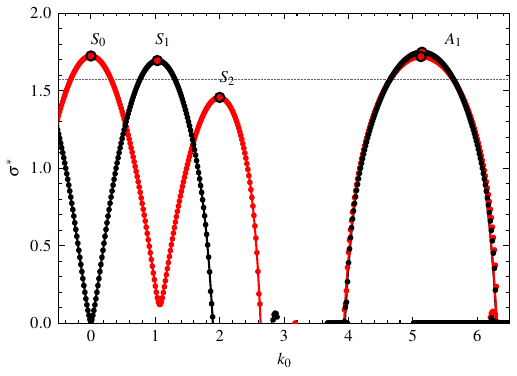}

  \caption{
    Stability curves for the case of two helical pairs {\it(a)} without and {\it
    (b)} with a central hub vortex and ($R^*=7$, $H^*/R^*=1.5$, $\beta=3/4$).
    Colour indicates the phase difference between the two helical pairs:
    $\varphi=0$ (in black), $\varphi=\pi$ (in red). For reference, fig {\it(a)}
    includes the values for two uniform helices ($H/R=1.5$, $a_e/R=0.1$) in
    dashed lines.
   }
  \label{newfig:16}
\end{figure}
%%%%%%%%%%%%%%%%%%%%%%%%%%%%%%%%%%%%%%%%%%%%%%%%%%%%%%%%%%%%%%%%%%%%%%%%%%%%%%%%
%%%%%%%%%%%%%%%%%%%%%%%%%%%%%%%%%%%%%%%%%%%%%%%%%%%%%%%%%%%%%%%%%%%%%%%%%%%%%%%%
\begin{figure}\centering\it

  (a) \hfill (b) \hfill ~ \\

  \includegraphics[width=0.495\linewidth, trim=250 0 125 90, clip]{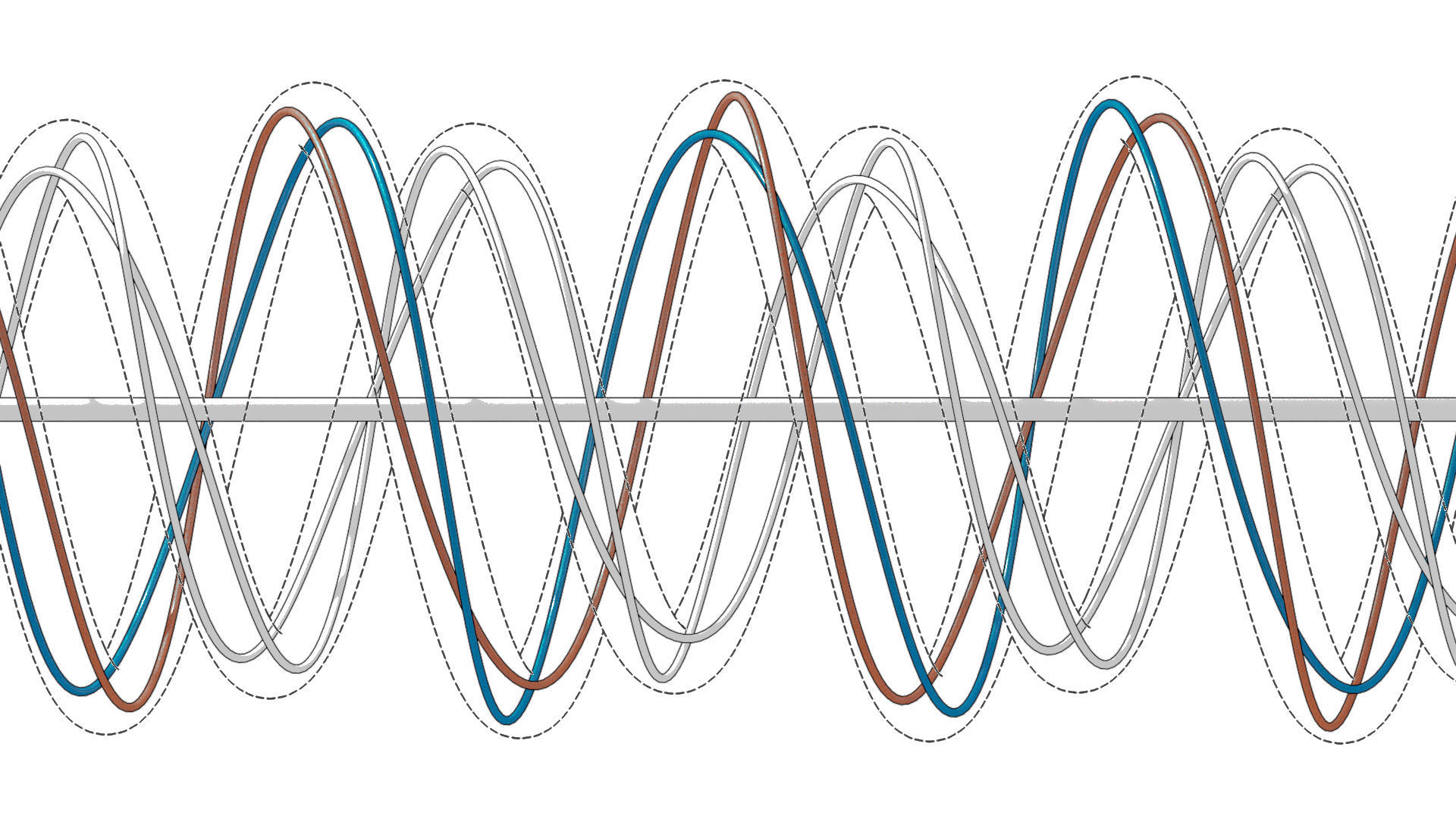}
  \includegraphics[width=0.495\linewidth, trim=0 0 0 0, clip]{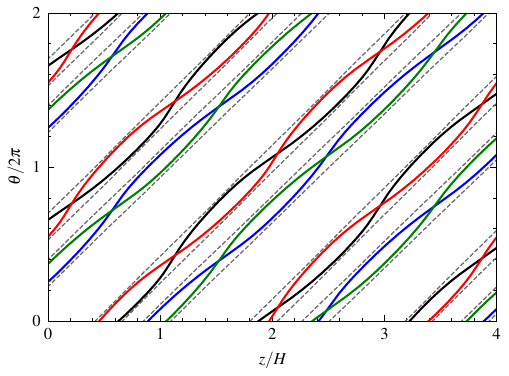}
  \caption{
    \emph{(a)} Three dimensional view of the base state from figure
    \ref{newfig:1}c perturbed by mode $S_0$, and \emph{(b)} developed plan view
    illustrating the global pairing.
  }
  \label{newfig:17}
\end{figure}
%%%%%%%%%%%%%%%%%%%%%%%%%%%%%%%%%%%%%%%%%%%%%%%%%%%%%%%%%%%%%%%%%%%%%%%%%%%%%%%%
\begin{table}\centering
  \begin{tabular}{crrcccc}
    Mode  & Frequency & \multicolumn{4}{c}{Wavenumbers} & Phase velocity\\
          & $\omega^*$& $k_0$    & $k_1$              & $k_2$             & $k_3$             & $c_0$   \\
    $S_0$ & $0.015$   & $ 0.000$ & $(-0.781, +0.719)$ & $(-1.531,+1.469)$ & $(-2.281,+2.219)$ & $0.952$ \\
    $S_1$ & $9.863$   & $-1.031$ & $(-1.781, +0.281)$ & $(-2.531,+0.469)$ & $(-3.281,+1.219)$ & $0.987$ \\
    $S_2$ & $19.086$  & $-2.000$ & $(-2.750, -1.250)$ & $(-3.500,-0.500)$ & $(-4.250,+0.250)$ & $0.984$ \\
    %$S_3$ & $27.663$  & $-2.906$ & $(-2.750, -1.250)$ & $(-3.500,-0.500)$ & $(-4.250,+0.250)$ & $0.984$ \\
    $A_1$ & $51.846$  & $-5.468$ & $(-6.219, -4.719)$ & $(-6.969,-3.969)$ & $(-7.719,-1.719)$ & $0.978$\\
    %$A_2$ & $51.442$  & $-5.546$ & $(-6.219, -4.719)$ & $(-6.969,-3.969)$ & $(-7.719,-1.719)$ & $0.978$
  \end{tabular}
  \caption{Leading wavenumbers of the modes shown in figure
  \ref{newfig:16}. Here, $\Delta k = 0.03125$.}
  \label{newtab:3}
\end{table}
%%%%%%%%%%%%%%%%%%%%%%%%%%%%%%%%%%%%%%%%%%%%%%%%%%%%%%%%%%%%%%%%%%%%%%%%%%%%%%%%

We consider similar geometric parameters as in \S\ref{sec4.1}, but increase the
pitch as to preserve the mean axial distance between neighbouring turns. The
results are summarized in figures \ref{newfig:16}, \ref{newfig:17} and
\ref{newfig:18}. The frequency spectra presented in figure \ref{newfig:16}a is
characterized by a first set of modes distributed over three overlapping lobes
at low frequencies, and a second set distributed over two additional lobes at
higher frequencies and a small lobe containing $A_0$. As in the previous case,
the former are symmetric modes, the latter are anti-symmetric. The structure of
the eigenvectors is the same as before with dominant wavenumbers at $k_0$ and
additional harmonic terms at $k_n=k_0 \pm n \beta$, see table \ref{newtab:3}. In
general, the maximum growth rate is larger than the maximum growth rate obtained
for two uniform helices with the total circulation of each vortex pair, which is
slightly larger than $\sigma = (\pi/2)(2\Gamma/H^2)$ obtained for an array of
point vortices of circulation $2\Gamma$ and separated by a distance $H/2$.

As expected, symmetric modes are reminiscent of the unstable modes obtained for
two equivalent helical vortices. Displacements between
neighbouring turns are out-of-phase for modes in the branch containing $S_0$ and
$S_2$ (in black), and phase-aligned for modes in the branches containing $S_1$ and
$S_3$ (in red). These displacements result in a local pairing at $2k$
azimuthal positions per turn of the large-scale helix. Mode $S_0$ corresponds to
a special case. Displacements between neighbouring turns are out-of-phase, where
one vortex pair expands in the radial direction, while the other one contracts,
resulting in a uniform pairing of the large-scale pattern along the azimuthal
direction (figures \ref{newfig:17}a-b). This is analogous to the global pairing
mode $k=0$ of the interlaced helices (see, for instance
\cite{okulov2010applications, quaranta2019local}).

Additionally, we note that anti-symmetric modes are observed over a similar range of
values as in the case of one vortex pair. One notable difference is that two
modes are now obtained for the same $k_0$ (each using a different colour in
figure \ref{newfig:16}a), where each pair has a similar structure but shifted in
phase (not shown).

{%
The introduction a central hub vortex ensures the total circulation to be zero
and the angular velocity to vanish as $r \to \infty$. This has a small effect on
the frame velocity and the base state with and without central hub are
qualitatively similar. The presence of a central hub modifies the stability
properties to a small degree (figures \ref{newfig:16}b). For instance, the case
with a central hub vortex has generally larger growth rates than the case
without by 2-3\%. Additionally, the branches containing $A_0$ are suppressed by
the hub vortex, while the out-of-phase perturbations are no longer neutrally
stable near $k_0=1$. In the following, we consider only the case with a central
hub.
}

\subsection{Geometric dependency of the most unstable modes}

%%%%%%%%%%%%%%%%%%%%%%%%%%%%%%%%%%%%%%%%%%%%%%%%%%%%%%%%%%%%%%%%%%%%%%%%%%%%%%%%
\begin{figure}\it
  (a) \hfill (b) \hfill (c) \hfill ~\\
  \includegraphics[width=\linewidth, trim=0 0 0 0, clip]{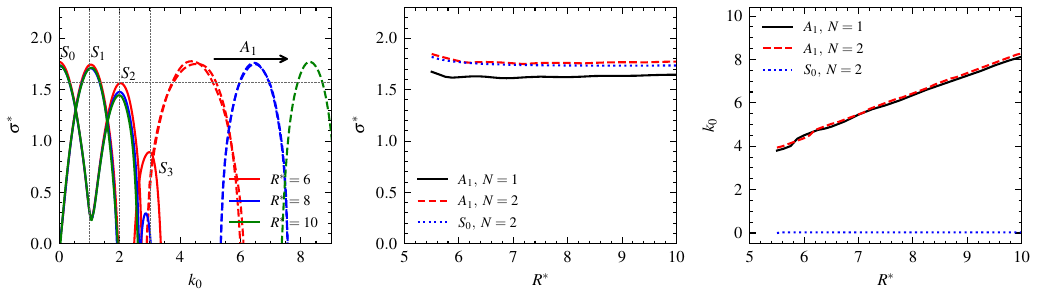}

  (d) \hfill (e) \hfill (f) \hfill ~\\
  \includegraphics[width=\linewidth, trim=0 0 0 0, clip]{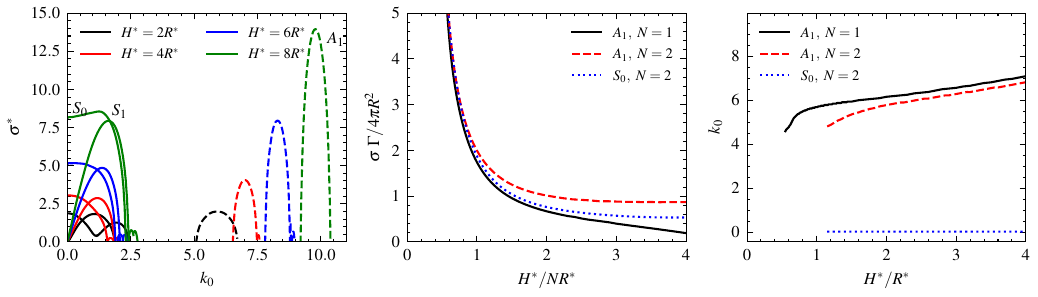}

  \caption{
  {\it (a,d)} Stability curves, {\it (b,e)} growth rates and {\it (c,f)}
  dominant wavenumbers as function of $R^*$ for $(H^*/R^*=1.5,\beta=3/4)$ {\it
  (a,b,c)}, and as function of $H^*/R^*$ for $(R^*=7,\beta=3/4)$ {\it
  (d,e,f)}.
  }
  \label{newfig:18}
\end{figure}
%%%%%%%%%%%%%%%%%%%%%%%%%%%%%%%%%%%%%%%%%%%%%%%%%%%%%%%%%%%%%%%%%%%%%%%%%%%%%%%%

As expected, symmetric and anti-symmetric modes display a different behaviour as
we vary the geometric parameters. Varying the separation distance has a small
influence on the symmetric modes. As in case of one vortex pair, the
change in growth rates is reminiscent to that of varying the effective core size
in uniform helices: small for $S_0$ and $S_1$, but more important for higher
wavenumbers (figure \ref{newfig:18}a). For anti-symmetric modes, the change in
growth rates is also small (figure \ref{newfig:18}b), while the dominant
wavenumber $k_0$ increases linearly with $R^*$ (figure \ref{newfig:18}c). For
$H^*/R^*=1.5$, $k_0$ was found to be roughly the same as in the case of one
vortex pair with equal effective pitch, suggesting the modes are selected
by the same pairing mechanism described in \S\ref{sec4}.

Varying the relative pitch shows the transition between two different regimes.
For small $H^*/R^*$, stability curves have a similar structure
as before: symmetric modes distributed over two or more branches with small
$k_0$ and anti-symmetric modes distributed over two overlapping branches with larger
$k_0$. Growth rates are larger than predicted rates for the equivalent uniform
helices, but still close to the point vortex prediction $\sigma \sim \Gamma
N^2/H^2$. For large $H^*/R^*$, the stability curves are
characterized by two branches at small $k_0$ containing modes $S_0$ and $S_1$,
and single branch containing $A_1$ at larger $k_0$. Modes $S_1$ and $A_1$ are
shifted towards larger wavenumbers. Instead of vanishing, the dimensionless
growth rate $\sigma^*$ proceeds to increase, pointing to a different scaling
law with $\sigma\sim\Gamma/R^2$ for large $H^*/R^*$ (figures
\ref{newfig:16}d-e). A similar transition is observed for the most unstable
wavenumber $k_0$: $k_0 \sim H^*/NR^*$ for small $H^*/R^*$, and $k_0 \sim H^*/R^*$
for large $H^*/R^*$ (figure \ref{newfig:18}f).

This change of regime can be understood as follows. For the case a single pair
($N=1$), the limit of large $H^*/R^*$ leads to a pair of parallel co-rotating
vortices which are known to be stable with respect to long wavelength
perturbations \citep{jimenez1975stability}. For the case of $N=2$ vortex pairs
with a central hub, the same limit provides a system composed of two co-rotating
pairs of vortices of circulation $\Gamma$ and one counter-rotating vortex of
circulation $-4\Gamma$ at the centre. For this configuration, the instability is
necessarily controlled by the distance between the vorticity centroids ($b=2R$
in our current notation), and the separation distance $d$. This configuration is
similar to four-vortex systems \citep{crouch1997instability, fabre2002optimal},
which are known to be unstable, with a maximum growth rate scaling with
$\Gamma/b^2$, and a most unstable wavenumber varying with $b$ and $d$.

\section{Discussion}
\label{sec6}

%%%%%%%%%%%%%%%%%%%%%%%%%%%%%%%%%%%%%%%%%%%%%%%%%%%%%%%%%%%%%%%%%%%%%%%%%%%%%%%%
\begin{figure}\it\centering
  \hspace{1cm}(a) \hfill ~ \\
  \includegraphics[width=0.8\linewidth, trim=0 50 0 50, clip]{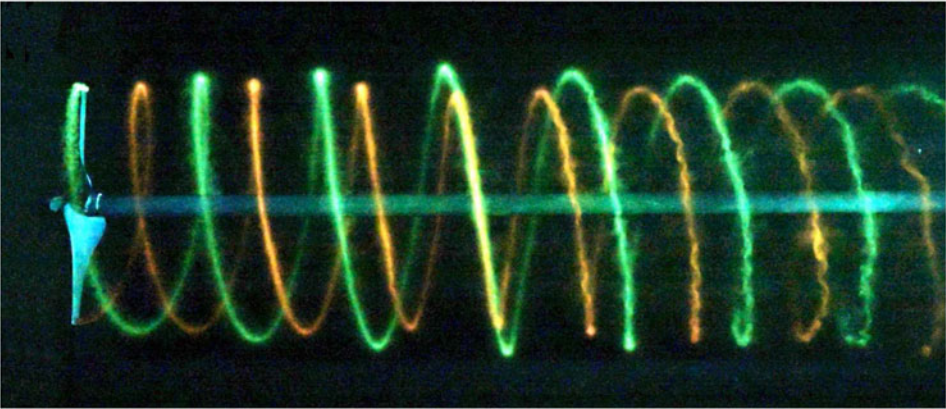}

  \hspace{1cm}(b) \hfill ~ \\
  \includegraphics[width=0.92\linewidth, trim=0 0 0 0, clip]{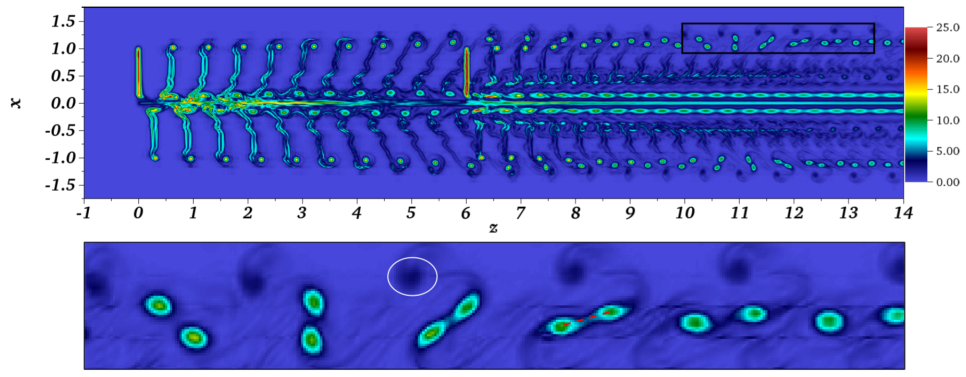}
  \caption{{%
  {\it (a)} Experimental dye visualisation for a two-bladed rotor with
   1.5\% radial rotor offset taken \cite{quaranta2019local}.
   {\it (b)} Vorticity contours for two in-line wind turbines from
   \cite{kleine2019tip}. A close-up of the vortex interaction is shown at the bottom.}
  }
  \label{newfig:23}
\end{figure}
%%%%%%%%%%%%%%%%%%%%%%%%%%%%%%%%%%%%%%%%%%%%%%%%%%%%%%%%%%%%%%%%%%%%%%%%%%%%%%%%

In this article, we have studied the long-wave stability properties of
closely-spaced helical vortex pairs using a cut-off filament approach. The
considered base flow configuration corresponds to the far-wake produced by a
rotor emitting two distinct vortices near the tips of each blade and which was
studied previously in \citep{castillo2020}. Both the temporal linear spectrum
and the linear impulse response have been analysed, but the linear spectrum has
been found to be much more convenient to identify the different instability
modes.

We have classified these modes in different groups. Symmetric modes are
characterized by a block displacement of the vortex pair, analogous to the local
pairing modes in helical geometries. Anti-symmetric modes are characterized by a
mirrored displacement with respect to the vorticity centre of the pair, with the
most unstable mode corresponding to a local pairing between one member of a pair
with the other member of a pair in the neighbouring turn. These modes are
particularly important since they could trigger the merging of vortex pairs.
{Additional modes are observed as we increase the twist parameter $\beta$: one
corresponds to a radial expansion of the pair and a displacement along the
centerline helix, while the other corresponds to the global and local pairing
modes of the vortex pair, analogous to the case of two interlaced helices
obtained by straightening the centerline helix.} We have also considered the
dependency of the stability properties with respect to the relative pitch, the
separation distance, and the twist parameter. We have identified the regions in
the parameter space where each mode is dominant. Our observations also suggest
that the pairing mechanism associated with the anti-symmetric modes amplifies a
specific axial wavelength (in the developed plane) instead of an azimuthal
wavelength, reminiscent of the anti-symmetric modes observed in four-vortex
systems. A similar pairing mechanism has also been observed for the case of two
pairs of helical vortices with one central hub. However, in this case, the
instability does not disappear in the limit of large pitch and exhibits a
maximum growth rate scaling with $R/d$. Additional, the central hub was found to
have only a small influence on the stability properties.

{
Experimental devices with 8cm and 24cm radius by \cite{schroder2020experimental,
schroder2021instability} successfully generated a pair of tip vortices. Their
main objective was to obtain a larger and less intense tip vortex. In their
case, tip vortices were unstable with respect to a centrifugal instability due
to patches of opposite signed vorticity remaining from the roll-up process. This
instability triggered the vortex merging long before the long-wave instabilities
could be observed. We speculate that it could be possible to delay such
instability by carefully tuning the blade geometry, or by considering a larger
rotor. Since core sizes typically scale with the chord length, very large rotors
could generate well separated tip vortices with stable cores. Because of the
vortex diffusion, merging is expected even in the absence of external
perturbations and would depend on the ratio between the core radii and the
separation distance. External perturbations would only accelerate this process.
Anti-symmetric modes are expected to trigger the merging faster than symmetric
modes. However, this would require a form of active control. Since,
anti-symmetric modes are excited by larger temporal and spatial frequencies,
it is possible they could be more easily excited by atmospheric turbulence than
symmetric modes.

Helical pairs are not necessarily limited to the case of a tip-splitting rotor.
One alternative to generate a helical pair would be to consider asymmetric
rotors. As shown in \cite{quaranta2019local}, a radial asymmetry excites the
global pairing mode to obtain a remarkably coherent structure like the one
displayed in figure \ref{newfig:23}a. From a topology perspective, this
structure is consistent with a leapfrogging wake with $d \sim H/2$ and $\beta\ll
1$, where the value of $\beta$ is controlled by the radial offset. This is
a promising approach since existing wind turbines could be easily modified.
A similar result can be obtain using an axial offset, as in the case of two
in-line wind turbines considered by \cite{kleine2019tip}. If the axial offset is
not too large and is a multiple of the pitch, we may expect the interaction
between tip vortices result in $N$ pairs of helical vortices downstream, like
the ones displayed in figure \ref{newfig:23}b. As observed in figure 5 of
\cite{kleine2019tip}, unstable dynamical modes are either in phase-alignment or
in phase-opposition, similar to the symmetric and anti-symmetric modes presented
here.
}

Our analysis assumes the wavelength of the displacement perturbations to be
large with respect to the core size. For helical vortices,
\cite{quaranta2015long} estimated the limit of validity in the form of
wavenumber $k \leq k_l$. For the values used in figure \ref{newfig:9}, this
upper limit corresponds to $k_l \approx 0.40(a_e/R)^{-1}$ and $k_l \approx
1.60(a_e/d)^{-1}$. Since we consider slender vortex filaments ($a_e/R\sim 0.01$ and
$a_e/d\sim 0.1$), the limit of validity of the long-wave approximation should
not be a concern. Unlike \cite{quaranta2015long}, which uses analytical
expressions, our approach considers the filaments as a sequence of straight
segments to compute the Jacobian matrix using semi-analytical expressions. This
approach has been validated using known results for the long-wave instability of
uniform helices. However, it is more general as it does not require prior
knowledge of the spatial structure of the instability modes. It also provides
the complete spectrum and applies to any stationary vortex solution, like the
ones in figure \ref{newfig:23}.

By using a filament approach, we have neglected what is occurring in the vortex
cores. Yet, vortex cores are expected to be distorted by curvature and straining
effects \citep{blanco2015internal}. Moreover, these deformations are also
responsible for the short-wave instabilities developing in vortex cores.
Depending on the geometric parameters, these short-wave instabilities
can become dominant. For instance, the elliptical instability is
expected to grow with $\Gamma/d^2$ instead of $\Gamma/H^2$ although with
different pre-factors \citep{roy2008stability, blanco2016elliptic}, while the
curvature instability \citep{blanco2017curvature} is also expected to be present
and important if the vortex core exhibits an axial jet.
None of these have been considered here. However, \cite{brynjell2020numerical}
have shown that it is indeed possible to  analyse the stability of uniform
helical vortices with respect to both short and large wavelengths using a DNS
from an initial condition obtained by the filament solutions together with a
prescribed vortex model in the cores.  It would be interesting to implement such
an approach to our configurations in order to analyse the competition between
both instabilities.

\backsection[Acknowledgements]{
The authors are grateful to
Eduardo Dur\'an Venegas and Thomas Leweke for their valuable contributions.
}

\backsection[Funding]{
This work is part of the French-German project TWIN-HELIX, supported by the
Agence Nationale de la Recherche (grant no. ANR-17-CE06-0018) and the Deutsche
Forschungsgemeinschaft (grant no. 391677260).
}

\backsection[Declaration of interests]{The authors report no conflict of interest.}

\backsection[Author ORCID]{
A. Castillo-Castellanos, https://orcid.org/0000-0003-2175-324X;
S. Le Dizès, https://orcid.org/0000-0001-6540-0433}

\appendix

\section{Space-time impulse response}
\label{secA}
\subsection{{Methodology}}
The space-time impulse response of $\boldsymbol{X}_j^B$ is studied as in
\cite{venegas2020}. We introduce a Dirac impulse as to excite all the wavenumber
components with equal amplitude and follow the evolution of the resulting
wavepacket. We consider two types of initial perturbation as to preferentially
excite the symmetric and anti-symmetric modes
\begin{equation}
  \boldsymbol{q}_0'(\zeta) =
  \begin{cases}
    (0,0,A_0 \delta(\zeta-\zeta_p),0,0,~~A_0 \delta(\zeta-\zeta_0)) & \mbox{Case A (in-phase)}
    \\
    (0,0,A_0 \delta(\zeta-\zeta_p),0,0,-A_0 \delta(\zeta-\zeta_0)) & \mbox{Case B (out-of-phase)}
  \end{cases}
\end{equation}
where $A_0$ and $\zeta_0$ indicate the amplitude and wake coordinate of the
initial perturbation and $\delta$ is the Dirac delta function.
Then, we use \eqref{sec2:eq16} recursively to obtain
$\boldsymbol{q}'(\zeta,\psi=m\Delta\psi)$ for $(m=1,2,\cdots)$ until the
exponential regime is established. In general, the exponential regime is
established quickly and our calculation domain is considered to be long enough
to avoid boundary effects.

Temporal growth rates are estimated from the impulse response as follows.
At each time $\psi$, we apply the spatial Fourier transform along $\zeta$ to
the axial displacement. Then, for each azimuthal wavenumber $k$ we can monitor
the amplitude of each Fourier mode and estimate its growth rate by
\begin{equation}\label{eq:sigma_lir}
  \sigma_j(k) \approx \frac{\partial (\log\lVert \hat{z}'_j(k)\rVert)}{\partial \psi}.
\end{equation}

It is also interesting to consider the growth rate
\begin{equation}\label{eq:def:sigmav}
  \sigma_V(V_\psi) \approx \frac{\log(\vert z'_j (\psi, \zeta_0 + V_\psi \psi) \vert)}{\psi}
\end{equation}
in the frame moving along the vortex structure with an angular velocity
$V_\psi$. In particular, we are interested in the velocity $V_\psi^{\max}$ at
which the growth rate is maximum and the upper and lower limits at which the
perturbation grows, $V_\psi^+$ and $V_\psi^-$. This provides a quantitative
criterium to identify convectively unstable and absolutely unstable flows
\citep{huerre1990local}.

\subsection{{Stability curves from the impulse response}}

%%%%%%%%%%%%%%%%%%%%%%%%%%%%%%%%%%%%%%%%%%%%%%%%%%%%%%%%%%%%%%%%%%%%%%%%%%%%%%%%
\begin{figure}\it\centering
  \hspace{0.5cm}(a) \hfill (b) \hfill ~ \\
  \includegraphics[width=0.95\linewidth, trim=0 0 0 0, clip]{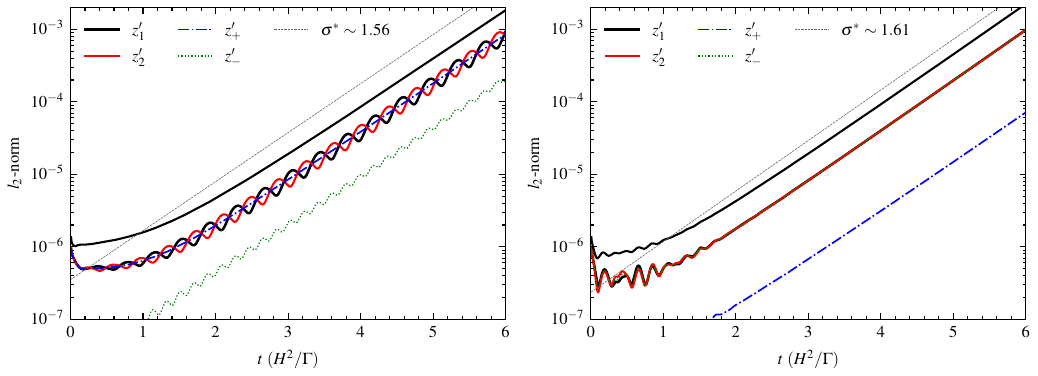}
  \caption{
    Linear impulse response for ($R^*=10$, $H^*/R^*=0.75$, $\beta=1$). {\it (a,b)}
    $l_2$-norm of the perturbations for cases A and B, respectively.
  }
  \label{fig:appendixa:3}
\end{figure}
%%%%%%%%%%%%%%%%%%%%%%%%%%%%%%%%%%%%%%%%%%%%%%%%%%%%%%%%%%%%%%%%%%%%%%%%%%%%%%%%

Consider the propagation of the initial perturbation corresponding to case A.
The norm of the total displacement $\boldsymbol{q}'$ grows exponentially with
constant rate $\sigma^*$ while the axial displacements $z_1'$ and $z_2'$ display
variable growth rates with a period comparable to the characteristic time of the
vortex pair (figure \ref{fig:appendixa:3}a). The symmetric component $z_+'$ is
dominant and displays a constant growth rate equal to $\sigma^*$. A similar
behaviour is observed for radial and angular displacements (not shown). The
spatio-temporal evolution corresponding to case B provides similar growth rates,
but $z_-'$ is now dominant and the oscillations in $z_1'$ and $z_2'$ are less
pronounced (figure \ref{fig:appendixa:3}b). Cases A and B have slightly
different growth rates, illustrating a clear dependency on the initial
conditions (figure \ref{fig:appendixa:3}a and \ref{fig:appendixa:3}b). The
observed growth rates correspond to the predicted rates for the most unstable
symmetric and anti-symmetric modes, $S_1$ and $A_1$, respectively. As expected,
if we observe long enough both cases eventually arrive to whichever one with the
largest growth rate.

%%%%%%%%%%%%%%%%%%%%%%%%%%%%%%%%%%%%%%%%%%%%%%%%%%%%%%%%%%%%%%%%%%%%%%%%%%%%%%%%
\begin{figure}\it\centering
  (a) \hfill (b) \hfill (c) \hfill ~ \\
  \includegraphics[width=\linewidth, trim=0 0 0 0, clip]{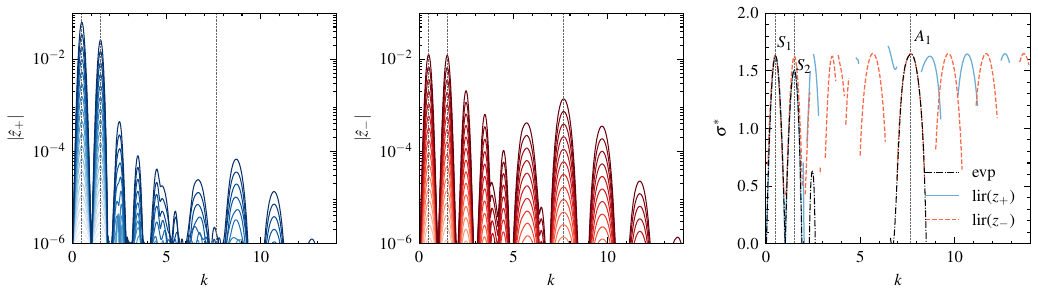}
  \caption{
    Fourier amplitudes of {\it (a)} $z_+'$ and {\it (b)} $z_-'$ for case A,
    where each line corresponds to a different instant at regular intervals, and
    {\it (c)} corresponding growth rates obtained using \eqref{eq:sigma_lir}
    compared to results from the eigenvalue problem.
  }
  \label{fig:appendixa:4}
\end{figure}
%%%%%%%%%%%%%%%%%%%%%%%%%%%%%%%%%%%%%%%%%%%%%%%%%%%%%%%%%%%%%%%%%%%%%%%%%%%%%%%%
%%%%%%%%%%%%%%%%%%%%%%%%%%%%%%%%%%%%%%%%%%%%%%%%%%%%%%%%%%%%%%%%%%%%%%%%%%%%%%%%
\begin{figure}\it
  \centering
  (a) \hfill (b) \hfill (c) \hfill ~ \\
  \includegraphics[width=\linewidth, trim=0 0 0 0, clip]{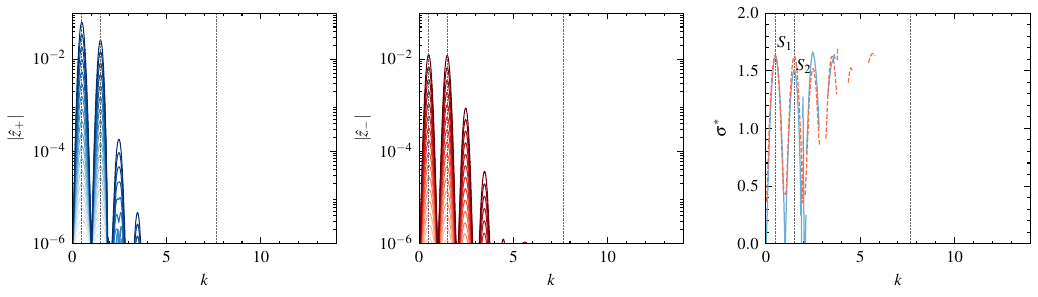}

  (d) \hfill (e) \hfill (f) \hfill ~ \\
  \includegraphics[width=\linewidth, trim=0 0 0 0, clip]{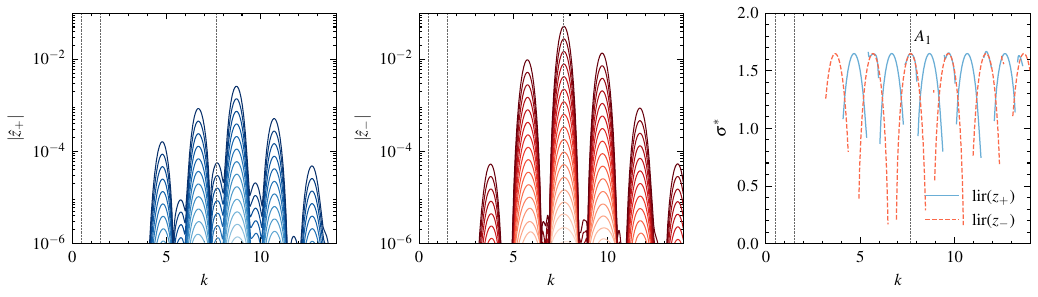}
  \caption{
    Fourier amplitudes of {\it (a,d)} $z_+'$ and {\it (b,e)} $z_-'$, where each
    line corresponds to a different instant, and {\it (c,d)} growth rates
    obtained using \eqref{eq:sigma_lir}. Figs. {\it (a-c)} correspond to case A
    with a high-pass filter, and figs. {\it (d-f)}
    to case B with a low-pass filter, see text.
  }
  \label{fig:appendixa:5}
\end{figure}
%%%%%%%%%%%%%%%%%%%%%%%%%%%%%%%%%%%%%%%%%%%%%%%%%%%%%%%%%%%%%%%%%%%%%%%%%%%%%%%%

Figure \ref{fig:appendixa:4}a (resp. \ref{fig:appendixa:4}b) displays the
Fourier spectra of $z_+'$ (resp. $z_-'$) taken at regular intervals, where the
most unstable wavenumbers are clearly identified. For each wavenumber $k$, the
corresponding growth rate obtained using \eqref{eq:sigma_lir} is shown in figure
\ref{fig:appendixa:4}c. Results are in good agreement with the eigenvalue
problem (dash-dotted lines). For instance, in $z_+'$ the most unstable
wavenumbers correspond to the dominant wavenumbers of $S_1$ and $S_2$.
Additional peaks in $z_+'$ (in blue) correspond to even harmonics, while peaks
in $z_-'$ (in orange) correspond to odd harmonics of $S_1$. Similar results are
observed for mode $A_1$ and to a smaller degree for mode $S_2$.

Issues in recovering the harmonics of mode $S_2$ can be explained by the
overlapping with other modes. We may take advantage of the separation in
temporal frequency to reduce this effect. For instance, applying a high-pass
filter to the wave-packets from case A before evaluating the Fourier
coefficients, i.e., filtering the anti-symmetric modes, we recover the growth
rates of symmetric modes over a wider range of wavenumbers (figures
\ref{fig:appendixa:5}a-c). Conversely, applying a low-pass filter to the
wave-packets from case B, i.e., filtering the symmetric modes, we recover the
growth rates of anti-symmetric modes over a similar range (figures
\ref{fig:appendixa:5}d-f). Superimposing the growth rates obtained from cases A
and B, we obtain a more complete picture of the stability curves, in good
agreement with results from section \ref{sec4}.

%%%%%%%%%%%%%%%%%%%%%%%%%%%%%%%%%%%%%%%%%%%%%%%%%%%%%%%%%%%%%%%%%%%%%%%%%%%%%%%%
\begin{figure}\it\centering
  (a) \hfill (b) \hfill (c) \hfill ~ \\
  \includegraphics[width=\linewidth, trim=0 0 0 0, clip]{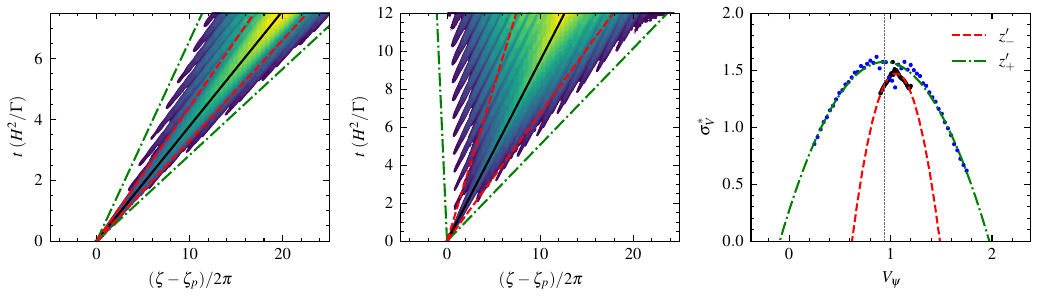}
  \caption{{
    Impulse response for ($R^*=10$, $\beta=1$) and case B: contours of
    $\log(\vert z'_1 \vert)$ for {\it (a)}  $H^*/R^*=0.75$ and {\it (b)}
    $H^*/R^*=0.5$ ; and {\it (c)} $\sigma_V^*$ as function of $V_{\psi}$ for the
    case in figure {\it (b)}. Here, $z'_+$ and $z'_-$, spread at different rates
    shown in green and red lines.}
  }
  \label{fig:appendixa:6}
\end{figure}
%%%%%%%%%%%%%%%%%%%%%%%%%%%%%%%%%%%%%%%%%%%%%%%%%%%%%%%%%%%%%%%%%%%%%%%%%%%%%%%%

\subsection{{Space-time evolution of perturbations}}

Figures \ref{fig:appendixa:6}a and \ref{fig:appendixa:6}b display the space-time
evolution of the initial perturbation corresponding to case B. In both cases,
perturbations initially propagate in a narrow wavepacket of high (temporal)
frequency content associated with anti-symmetric modes, before a second wavepacket
with lower frequencies typical of symmetric modes is observed. The case in figure
\ref{fig:appendixa:6}a corresponds to a convective instability, while the one
in figure \ref{fig:appendixa:6}b illustrates the transition to an absolute
instability.
Despite the overlapping of frequencies and wavenumbers, it is still possible to
estimate the front velocities using \eqref{eq:def:sigmav} by considering
separately the symmetric and anti-symmetric parts, $z'_+$ and $z'_-$. The
maximum growth rate corresponds to a velocity $V_\psi=V_\psi^{\max}$ close to
the phase velocity $c_0$ of the most unstable modes (see, peak value in figure
\ref{fig:appendixa:6}c and black lines in figures \ref{fig:appendixa:6}a-b). In
general, the anti-symmetric wavepacket (in dashed red lines) spreads more slowly
than symmetric part (in dash-dotted green lines), suggesting that the transition
from convective to absolute instability can be monitored by following the spread
of the symmetric part. As expected, the growth rate $\sigma_V^*$ of the
symmetric part behaves similarly to the equivalent helical vortex and
is well approximated by the predicted growth rate for a periodic
array of point vortices as proposed by \cite{venegas2020}
\begin{equation}\label{eq:appendixA:sigma_rel}
\sigma_V^* = (\pi/2)(1-(V_{rel}^*)^2) \end{equation} where
$V_{rel}^*=(V_\psi-V_\psi^{\max}) (2\pi/\Omega_R)/ t_{adv} $ is the frame velocity
relative to the advection frame.

\bibliographystyle{jfm}
\bibliography{article2}

\begin{thebibliography}{47}
\expandafter\ifx\csname natexlab\endcsname\relax\def\natexlab#1{#1}\fi
\def\au#1{#1} \def\ed#1{#1} \def\yr#1{#1}\def\at#1{#1}\def\jt#1{\textit{#1}}
  \def\bt#1{#1}\def\bvol#1{\textbf{#1}} \def\vol#1{#1} \def\pg#1{#1}
  \def\publ#1{#1}\def\arxiv#1{#1}\def\org#1{#1}\def\st#1{\textit{#1}}

\bibitem[Bayly(1988)]{bayly1988three}
{\sc \au{Bayly, B.~J.}} \yr{1988}  \at{Three-dimensional centrifugal-type
  instabilities in inviscid two-dimensional flows}.  \jt{Phys. Fluids}
  \bvol{31}~(1),  \pg{56--64}.

\bibitem[Bhagwat \& Leishman(2000)]{Bhagwat00}
{\sc \au{Bhagwat, M.~J.} \& \au{Leishman, J.~G.}} \yr{2000}  \at{Stability
  analysis of helicopter rotor wakes in axial flight}.  \jt{J. Amer. Helic.
  Soc.}  \bvol{45}~(3),  \pg{165}.

\bibitem[Blanco-Rodr{\'{\i}}guez \& Le~Diz{\`{e}}s(2016)]{blanco2016elliptic}
{\sc \au{Blanco-Rodr{\'{\i}}guez, F.~J.} \& \au{Le~Diz{\`{e}}s, S.}} \yr{2016}
  \at{Elliptic instability of a curved batchelor vortex}.  \jt{J. Fluid Mech.}
  \bvol{804},  \pg{224--247}.

\bibitem[Blanco-Rodr{\'{\i}}guez \& Le~Diz{\`{e}}s(2017)]{blanco2017curvature}
{\sc \au{Blanco-Rodr{\'{\i}}guez, F.~J.} \& \au{Le~Diz{\`{e}}s, S.}} \yr{2017}
  \at{Curvature instability of a curved batchelor~vortex}.  \jt{J. Fluid Mech.}
   \bvol{814},  \pg{397--415}.

\bibitem[Blanco-Rodr{\'\i}guez {\em et~al.\/}(2015)Blanco-Rodr{\'\i}guez,
  Le~Diz{\`e}s, Sel{\c{c}}uk, Delbende \& Rossi]{blanco2015internal}
{\sc \au{Blanco-Rodr{\'\i}guez, F.~J.}, \au{Le~Diz{\`e}s, S.},
  \au{Sel{\c{c}}uk, C.}, \au{Delbende, I.} \& \au{Rossi, M.}} \yr{2015}
  \at{{Internal structure of vortex rings and helical vortices}}.  \jt{J. Fluid
  Mech.}  \bvol{785},  \pg{219--247}.

\bibitem[Brocklehurst \& Pike(1994)]{brocklehurst1994reduction}
{\sc \au{Brocklehurst, A.} \& \au{Pike, A.C.}} \yr{1994} {Reduction of BVI
  noise using a vane tip}.  \bt{In {\em AHS Aeromechanics Specialists
  Conference\/}}.  \publ{American Helicopter Society}.

\bibitem[Brown {\em et~al.\/}(2022)Brown, Houck, Maniaci, Westergaard \&
  Kelley]{brown2022accelerated}
{\sc \au{Brown, K.}, \au{Houck, D.}, \au{Maniaci, D.}, \au{Westergaard, C.} \&
  \au{Kelley, C.}} \yr{2022}  \at{Accelerated wind-turbine wake recovery
  through actuation of the tip-vortex instability}.  \jt{AIAA Journal}  \pg{pp.
  1--13}.

\bibitem[Brynjell-Rahkola \& Henningson(2020)]{brynjell2020numerical}
{\sc \au{Brynjell-Rahkola, M.} \& \au{Henningson, D.~S.}} \yr{2020}
  \at{{Numerical realization of helical vortices: application to vortex
  instability}}.  \jt{Theor. Comp. Fluid Dyn}  \bvol{34}~(1),  \pg{1--20}.

\bibitem[Ca{\~n}adillas {\em et~al.\/}(2020)Ca{\~n}adillas, Foreman, Barth,
  Siedersleben, Lampert, Platis, Djath, Schulz-Stellenfleth, Bange, Emeis {\em
  et~al.\/}]{canadillas2020offshore}
{\sc \au{Ca{\~n}adillas, B.}, \au{Foreman, R.}, \au{Barth, V.},
  \au{Siedersleben, S.}, \au{Lampert, A.}, \au{Platis, A.}, \au{Djath, B.},
  \au{Schulz-Stellenfleth, J.}, \au{Bange, J.}, \au{Emeis, S.} \& \au{others}}
  \yr{2020}  \at{Offshore wind farm wake recovery: Airborne measurements and
  its representation in engineering models}.  \jt{Wind Energy}  \bvol{23}~(5),
  \pg{1249--1265}.

\bibitem[Castillo-Castellanos {\em et~al.\/}(2021)Castillo-Castellanos,
  Le~Diz\`es \& Dur\'an~Venegas]{castillo2020}
{\sc \au{Castillo-Castellanos, A.}, \au{Le~Diz\`es, S.} \& \au{Dur\'an~Venegas,
  E.}} \yr{2021}  \at{Closely spaced corotating helical vortices: General
  solutions}.  \jt{Phys. Rev. Fluids}  \bvol{6},  \pg{114701}.

\bibitem[Crouch(1997)]{crouch1997instability}
{\sc \au{Crouch, J.D.}} \yr{1997}  \at{Instability and transient growth for two
  trailing-vortex pairs}.  \jt{J. Fluid Mech.}  \bvol{350},  \pg{311--330}.

\bibitem[Crow(1970)]{crow1970stability}
{\sc \au{Crow, S.~C.}} \yr{1970}  \at{Stability theory for a pair of trailing
  vortices}.  \jt{AIAA journal}  \bvol{8}~(12),  \pg{2172--2179}.

\bibitem[Dur{\'a}n~Venegas \& Le~Diz{\`e}s(2019)]{venegas2019generalized}
{\sc \au{Dur{\'a}n~Venegas, E.} \& \au{Le~Diz{\`e}s, S.}} \yr{2019}
  \at{{Generalized helical vortex pairs}}.  \jt{J. Fluid Mech.}  \bvol{865},
  \pg{523--545}.

\bibitem[Dur{\'a}n~Venegas {\em et~al.\/}(2021)Dur{\'a}n~Venegas, Rieu \&
  Le~Diz{\`e}s]{venegas2020}
{\sc \au{Dur{\'a}n~Venegas, E.}, \au{Rieu, P.} \& \au{Le~Diz{\`e}s, S.}}
  \yr{2021}  \at{{Structure and stability of Joukowski’s rotor wake model}}.
  \jt{J. Fluid Mech.}  \bvol{911},  \pg{A6}.

\bibitem[Fabre \& Jacquin(2000)]{fabre2000stability}
{\sc \au{Fabre, D.} \& \au{Jacquin, L.}} \yr{2000}  \at{Stability of a
  four-vortex aircraft wake model}.  \jt{Phys. Fluids}  \bvol{12}~(10),
  \pg{2438--2443}.

\bibitem[Fabre {\em et~al.\/}(2002)Fabre, Jacquin \& Loof]{fabre2002optimal}
{\sc \au{Fabre, D.}, \au{Jacquin, L.} \& \au{Loof, A.}} \yr{2002}  \at{Optimal
  perturbations in a four-vortex aircraft wake in counter-rotating
  configuration}.  \jt{J. Fluid Mech.}  \bvol{451},  \pg{319}.

\bibitem[Frederik {\em et~al.\/}(2020)Frederik, Doekemeijer, Mulders \& van
  Wingerden]{frederik2020helix}
{\sc \au{Frederik, J.~A.}, \au{Doekemeijer, B.~M.}, \au{Mulders, S.~P.} \&
  \au{van Wingerden, J.-W.}} \yr{2020}  \at{The helix approach: Using dynamic
  individual pitch control to enhance wake mixing in wind farms}.  \jt{Wind
  Energy}  \bvol{23}~(8),  \pg{1739--1751}.

\bibitem[Fukumoto \& Miyazaki(1991)]{Fukumoto1991}
{\sc \au{Fukumoto, Y.} \& \au{Miyazaki, T.}} \yr{1991}  \at{Three--dimensional
  distorsions of a vortex filament zith axial velocity}.  \jt{J. Fluid Mech.}
  \bvol{222},  \pg{369--416}.

\bibitem[Gupta \& Loewy(1974)]{gupta1974theoretical}
{\sc \au{Gupta, B.P.} \& \au{Loewy, R.G.}} \yr{1974}  \at{Theoretical analysis
  of the aerodynamic stability of multiple, interdigitated helical vortices}.
  \jt{AIAA J.}  \bvol{12}~(10),  \pg{1381--1387}.

\bibitem[Hardin(1982)]{hardin1982velocity}
{\sc \au{Hardin, J.~C.}} \yr{1982}  \at{{The velocity field induced by a
  helical vortex filament}}.  \jt{Phys. Fluids}  \bvol{25}~(11),
  \pg{1949--1952}.

\bibitem[Huang {\em et~al.\/}(2019)Huang, Moghadam, Meysonnat, Meinke \&
  Schr{\"o}der]{huang2019numerical}
{\sc \au{Huang, X.}, \au{Moghadam, S. M.~A.}, \au{Meysonnat, P.S.}, \au{Meinke,
  M.} \& \au{Schr{\"o}der, W.}} \yr{2019}  \at{Numerical analysis of the effect
  of flaps on the tip vortex of a wind turbine blade}.  \jt{Int. J. Heat Fluid
  Flow}  \bvol{77},  \pg{336--351}.

\bibitem[Huerre \& Monkewitz(1990)]{huerre1990local}
{\sc \au{Huerre, P.} \& \au{Monkewitz, P.~A.}} \yr{1990}  \at{Local and global
  instabilities in spatially developing flows}.  \jt{Annu. Rev. Fluid Mech.}
  \bvol{22}~(1),  \pg{473--537}.

\bibitem[Ivanell {\em et~al.\/}(2010)Ivanell, Mikkelsen, S{\o{}}rensen \&
  Henningson]{ivanell2010stability}
{\sc \au{Ivanell, S.}, \au{Mikkelsen, R.}, \au{S{\o{}}rensen, J.~N.} \&
  \au{Henningson, D.}} \yr{2010}  \at{Stability analysis of the tip vortices of
  a wind turbine}.  \jt{Wind Energy}  \bvol{13}~(8),  \pg{705--715}.

\bibitem[Jimenez(1975)]{jimenez1975stability}
{\sc \au{Jimenez, J.}} \yr{1975}  \at{Stability of a pair of co-rotating
  vortices}.  \jt{Phys. Fluids}  \bvol{18}~(11),  \pg{1580--1581}.

\bibitem[Josserand \& Rossi(2007)]{josserand2007merging}
{\sc \au{Josserand, Ch.} \& \au{Rossi, M.}} \yr{2007}  \at{The merging of two
  co-rotating vortices: a numerical study}.  \jt{European Journal of
  Mechanics-B/Fluids}  \bvol{26}~(6),  \pg{779--794}.

\bibitem[Kawada(1936)]{kawada1936induced}
{\sc \au{Kawada, S.}} \yr{1936}  \at{Induced velocity by helical vortices}.
  \jt{J. Aeronaut. Sci.}  \bvol{3}~(3),  \pg{86--87}.

\bibitem[Kerswell(2002)]{kerswell2002elliptical}
{\sc \au{Kerswell, R.~R.}} \yr{2002}  \at{Elliptical instability}.  \jt{Annu.
  Rev. Fluid Mech.}  \bvol{34}~(1),  \pg{83--113}.

\bibitem[Kleine {\em et~al.\/}(2019)Kleine, Kleusberg, Hanifi \&
  Henningson]{kleine2019tip}
{\sc \au{Kleine, V.G.}, \au{Kleusberg, E.}, \au{Hanifi, A.} \& \au{Henningson,
  D.S.}} \yr{2019} Tip-vortex instabilities of two in-line wind turbines.
  \bt{In {\em Journal of Physics: Conference Series\/}}, ,  \vol{vol. 1256},
  \pg{p. 012015}. IOP Publishing.

\bibitem[Lamb(1945)]{lamb1945hydrodynamics}
{\sc \au{Lamb, H.}} \yr{1945} {\em Hydrodynamics\/}.  \publ{Dover
  publications}.

\bibitem[Leishman {\em et~al.\/}(2002)Leishman, Bhagwat \&
  Bagai]{leishman2002free}
{\sc \au{Leishman, J.~G.}, \au{Bhagwat, M.~J.} \& \au{Bagai, A.}} \yr{2002}
  \at{Free-vortex filament methods for the analysis of helicopter rotor wakes}.
   \jt{J Aicr}  \bvol{39}~(5),  \pg{759--775}.

\bibitem[Levy \& Forsdyke(1928)]{Levy1928}
{\sc \au{Levy, H.} \& \au{Forsdyke, A.~G.}} \yr{1928}  \at{The steady motion
  and stability of a helical vortex}.  \jt{Proc. R. Soc. Lond. {\em A}}
  \bvol{120},  \pg{670--690}.

\bibitem[Leweke {\em et~al.\/}(2016)Leweke, Le~Dizes \&
  Williamson]{leweke2016dynamics}
{\sc \au{Leweke, T.}, \au{Le~Dizes, S.} \& \au{Williamson, C.~H.K.}} \yr{2016}
  \at{Dynamics and instabilities of vortex pairs}.  \jt{Annu. Rev. Fluid Mech.}
   \bvol{48}~(1),  \pg{507--541}.

\bibitem[Leweke {\em et~al.\/}(2014)Leweke, Quaranta, Bolnot,
  Blanco-Rodri\'{i}guez \& {Le Diz\`es}]{Leweke14}
{\sc \au{Leweke, T.}, \au{Quaranta, H.~U.}, \au{Bolnot, H.},
  \au{Blanco-Rodri\'{i}guez, F.~J.} \& \au{{Le Diz\`es}, S.}} \yr{2014}
  \at{Long- and short-wave instabilities in helical vortices}.  \jt{J. Phys.:
  Conf. Ser.}  \bvol{524},  \pg{012154}.

\bibitem[Meunier {\em et~al.\/}(2002)Meunier, Ehrenstein, Leweke \&
  Rossi]{meunier2002merging}
{\sc \au{Meunier, P.}, \au{Ehrenstein, U.}, \au{Leweke, T.} \& \au{Rossi, M.}}
  \yr{2002}  \at{A merging criterion for two-dimensional co-rotating vortices}.
   \jt{Phys. Fluids}  \bvol{14}~(8),  \pg{2757--2766}.

\bibitem[Meunier \& Leweke(2005)]{meunier2005elliptic}
{\sc \au{Meunier, P.} \& \au{Leweke, T.}} \yr{2005}  \at{Elliptic instability
  of a co-rotating vortex pair}.  \jt{J. Fluid Mech.}  \bvol{533},
  \pg{125--159}.

\bibitem[Okulov(2004)]{okulov2004stability}
{\sc \au{Okulov, V.~L.}} \yr{2004}  \at{On the stability of multiple helical
  vortices}.  \jt{J. Fluid Mech.}  \bvol{521},  \pg{319--342}.

\bibitem[Okulov \& S{\o{}}rensen(2007)]{Okulov07}
{\sc \au{Okulov, V.~L.} \& \au{S{\o{}}rensen, J.~N.}} \yr{2007}  \at{Stability
  of helical tip vortices in a rotor far wake}.  \jt{J. Fluid Mech.}
  \bvol{576},  \pg{1--25}.

\bibitem[Okulov \& S{\o{}}rensen(2009)]{okulov2010applications}
{\sc \au{Okulov, V.~L.} \& \au{S{\o{}}rensen, J.~N.}} \yr{2009}
  \at{Applications of 2d helical vortex dynamics}.  \jt{Theor. Comp. Fluid Dyn}
   \bvol{24}~(1-4),  \pg{395--401}.

\bibitem[Quaranta {\em et~al.\/}(2015)Quaranta, Bolnot \&
  Leweke]{quaranta2015long}
{\sc \au{Quaranta, H.~U.}, \au{Bolnot, H.} \& \au{Leweke, T.}} \yr{2015}
  \at{Long-wave instability of a helical vortex}.  \jt{J. Fluid Mech.}
  \bvol{780},  \pg{687--716}.

\bibitem[Quaranta {\em et~al.\/}(2019)Quaranta, Brynjell-Rahkola, Leweke \&
  Henningson]{quaranta2019local}
{\sc \au{Quaranta, H.~U.}, \au{Brynjell-Rahkola, M.}, \au{Leweke, T.} \&
  \au{Henningson, D.~S.}} \yr{2019}  \at{Local and global pairing instabilities
  of two interlaced helical vortices}.  \jt{J. Fluid Mech.}  \bvol{863},
  \pg{927--955}.

\bibitem[Robinson \& Saffman(1982)]{Robinson1982}
{\sc \au{Robinson, A.~C.} \& \au{Saffman, P.~G.}} \yr{1982}
  \at{Three-dimensional stability of vortex arrays}.  \jt{J. Fluid Mech.}
  \bvol{125},  \pg{411}.

\bibitem[Roy {\em et~al.\/}(2008)Roy, Schaeffer, Le~Diz{\`{e}}s \&
  Thompson]{roy2008stability}
{\sc \au{Roy, C.}, \au{Schaeffer, N.}, \au{Le~Diz{\`{e}}s, S.} \& \au{Thompson,
  M.}} \yr{2008}  \at{Stability of a pair of co-rotating vortices with axial
  flow}.  \jt{Phys. Fluids}  \bvol{20}~(9),  \pg{094101}.

\bibitem[Schr{\"o}der {\em et~al.\/}(2020)Schr{\"o}der, Leweke,
  H{\"o}rnschemeyer \& Stumpf]{schroder2020experimental}
{\sc \au{Schr{\"o}der, D.}, \au{Leweke, T.}, \au{H{\"o}rnschemeyer, R.} \&
  \au{Stumpf, E.}} \yr{2020} Experimental investigation of a helical vortex
  pair.  \bt{In {\em Deutscher Luft-und Raumfahrtkongress 2020\/}}.

\bibitem[Schr{\"o}der {\em et~al.\/}(2021)Schr{\"o}der, Leweke,
  H{\"o}rnschemeyer \& Stumpf]{schroder2021instability}
{\sc \au{Schr{\"o}der, D.}, \au{Leweke, T.}, \au{H{\"o}rnschemeyer, R.} \&
  \au{Stumpf, E.}} \yr{2021}  \at{Instability and merging of a helical vortex
  pair in the wake of a rotor}.  \jt{J. Phys.: Conf. Ser.}  \bvol{1934}~(1),
  \pg{012007}.

\bibitem[Sel{\c{c}}uk {\em et~al.\/}(2017)Sel{\c{c}}uk, Delbende \&
  Rossi]{selccuk2017helical}
{\sc \au{Sel{\c{c}}uk, C.}, \au{Delbende, I.} \& \au{Rossi, M.}} \yr{2017}
  \at{Helical vortices: Quasiequilibrium states and their time evolution}.
  \jt{Phys. Rev. Fluids}  \bvol{2}~(8),  \pg{084701}.

\bibitem[Walther {\em et~al.\/}(2007)Walther, Gu\'enot, Machefaux, Rasmussen,
  Chatelain, Okulov, S{\o{}}rensen, Bergdorf \& Koumoutsakos]{Walther07}
{\sc \au{Walther, J.~H.}, \au{Gu\'enot, M.}, \au{Machefaux, E.}, \au{Rasmussen,
  J.~T.}, \au{Chatelain, P.}, \au{Okulov, V.~L.}, \au{S{\o{}}rensen, J.~N.},
  \au{Bergdorf, M.} \& \au{Koumoutsakos, P.}} \yr{2007}  \at{A numerical study
  of the stabilitiy of helical vortices using vortex methods}.  \jt{J. Phys.:
  Conf. Ser.}  \bvol{75},  \pg{012034}.

\bibitem[Widnall(1972)]{widnall1972stability}
{\sc \au{Widnall, S.~E.}} \yr{1972}  \at{The stability of a helical vortex
  filament}.  \jt{J. Fluid Mech.}  \bvol{54}~(4),  \pg{641--663}.

\end{thebibliography}

\end{document}